\documentclass[10pt]{artikel3}
\usepackage{amsfonts,amsmath,amssymb,marvosym}

\usepackage[dvips]{feynmp}

\usepackage{epsfig,slashed}

\usepackage{float,braket}
\usepackage{subfigure,pslatex}

\usepackage{graphicx,wrapfig,color}

\newcommand{\pp}{\ensuremath{p\!\!\!\!/}}

\newcommand{\p}{\partial}
\newcommand{\oc}{\overline{c}}

\newcommand{\omu}{\overline{\mu}}

\renewcommand{\d}{\ensuremath{\mathrm{d}}}

\newcommand{\opsi}{\overline{\psi}}

\DeclareSymbolFont{extraup}{U}{zavm}{m}{n}
\DeclareMathSymbol{\varheart}{\mathalpha}{extraup}{86}
\DeclareMathSymbol{\vardiamond}{\mathalpha}{extraup}{87}

\newcommand{\D}{\ensuremath{D\!\!\!\!/}}
\definecolor{Rood}{rgb}{1, 0, 0} % Red of svgnames

\begin{document}
\title{\noindent {\bf Confinement and dynamical chiral symmetry breaking in a non-perturbative renormalizable quark model }}
\author{D.~Dudal$^{a,b}$\thanks{david.dudal@kuleuven-kulak.be},\; M.S.~Guimaraes$^{c}$\thanks{msguimaraes@uerj.br},\; L.F.~Palhares$^{c,d}$\thanks{l.palhares@thphys.uni-heidelberg.de},\; S.P.~Sorella$^{c}$\thanks{sorella@uerj.br}\\\\
{\small  \textnormal{$^{a}$ Department of Physics, KU Leuven Campus Kortrijk - KULAK, Etienne Sabbelaan 53, 8500 Kortrijk, Belgium }}
\\
\small \textnormal{$^{b}$ Ghent University, Department of Physics and Astronomy, Krijgslaan 281-S9, 9000 Gent, Belgium }\\
\small \textnormal{$^{c}$ Departamento de F\'{\i }sica Te\'{o}rica, Instituto de F\'{\i }sica, UERJ - Universidade do Estado do Rio de Janeiro}\\
\small \textnormal{$^{d}$ Institut f\"ur Theoretische Physik, Universit\"at Heidelberg, Philosophenweg 16, 69120 Heidelberg, Germany}
 \normalsize}

\date{}
\maketitle
\begin{abstract}
Inspired by the construction of the Gribov-Zwanziger action in the Landau gauge, we introduce a quark model exhibiting
both confinement and chiral symmetry aspects. An important feature is the incorporation of spontaneous chiral symmetry breaking in a renormalizable fashion. The quark propagator in the condensed vacuum turns out to be of a confining type. Besides a real pole, it exhibits  complex conjugate poles. The resulting spectral form is explicitly shown to violate positivity,  indicative of its  unphysical character.
Moreover, the ensuing quark mass function fits  well to existing lattice data. To further validate the physical nature of the model, we  identify a massless pseudoscalar (i.e.~a pion) in the chiral limit and present  estimates for the $\rho$ meson mass  and decay constant.
\end{abstract}

\setcounter{page}{1}

\section{Introduction}
Next to confinement, i.e.~the absence of color charged particles in the QCD spectrum, the other crucial non-perturbative ingredient of QCD is the dynamical breaking of the chiral symmetry (D$\chi$SB). The latter is what ensures most of the meson/baryon mass and what explains the gap between the light pseudoscalar mesons and the rest of the spectrum due to their Goldstone boson nature. Albeit confinement and D$\chi$SB are well appreciated, their treatment still inspires active research. Even fundamental issues linger interesting, such as the nature of nuclear matter at large $N$ \cite{McLerran:2007qj} and the fact that in this limit chiral symmetry may be inhomogeneously broken \cite{Kojo:2009ha} and the possibility of an intimate connection between these two completely different phenomena suggested by the proximity of the corresponding thermal phase transitions observed on the lattice
\cite{Creutz:2011hy,Aoki:2006br}.

It turns out to be particularly difficult to have a model that incorporates  both features dynamically. Several efforts have been undertaken in the last years and in different directions. A few examples range from
Dyson-Schwinger equations \cite{Bashir:2013zha}, Hamiltonian models
 \cite{Fontoura:2012mz}, holographic approaches \cite{Li:2012ay} and phenomenologically based models with massive gluons \cite{Cornwall:2010ap}
   to Polyakov-extended Nambu-Jona-Lasinio (NJL) \cite{Fukushima:2003fw} or Quark-Meson (QM) \cite{Mocsy:2004yt,Megias:2004hj,Schaefer:2007pw} models as well as their extensions with nonlocal interactions \cite{Contrera:2010kz}
and  running analyses \cite{Hell:2008cc,Herbst:2010rf}.
In particular, extensions of chiral models that aim at incorporating confinement through Polyakov loops usually hinder on the unsolved problem of defining a kinetic term for these highly nonlocal objects. Moreover, if one considers strictly the spectrum of these Polyakov-extended quark models, the presence of asymptotic quarks is usually unavoidable, obscuring the connection with confinement.  The quark degrees of freedom described in those effective models appear to be different from the ones observed in the infrared regime of lattice QCD simulations which show that  the quark mass function is momentum dependent, attaining a non-vanishing value at the origin  in the chiral limit \cite{Parappilly:2005ei,Furui:2006ks,Burgio:2012ph}. Suitable non-local form factors, inspired by confinement, can be added to model non-perturbative quark propagators into NJL models  (cf.~e.g.~\cite{Bowler:1994ir,Plant:1997jr,Contrera:2007wu,
Loewe:2013zaa,Marquez:2014kla,Marquez:2015bca}) and to argue the suppression of free quarks \cite{Buballa:1992sz}.

One possible complementary way to face the  issue of confinement relies on  the Gribov approach \cite{Gribov:1977wm}, amounting to taking into account the non-perturbative effects of the existence of  Gribov copies which deeply affect the infrared dynamics of non-Abelian Yang-Mills theories.  This has resulted in a local and renormalizable action, known as the Gribov-Zwanziger action \cite{Vandersickel:2012tz},  and its more recent refined version \cite{Dudal:2007cw}. This set up turns out to be useful in order to capture aspects of gluon confinement. In fact, taking the example of the Landau gauge, a calculational framework exists for which a set of complex conjugate poles emerges in the gluon propagator, whose  tree level expression obtained from the Refined Gribov-Zwanziger action contains dynamical mass parameters $(\hat M^2$, $\hat m^2, \lambda^4)$, being given by
\begin{equation}\label{rgzgll}
  D(p^2)=\frac{p^2+\hat M^2}{p^4+(\hat M^2+\hat m^2)p^2+\lambda^4}  \;.
\end{equation}
Remarkably, expression \eqref{rgzgll} fits in a very accurate way the most recent lattice data on the gluon propagator \cite{Dudal:2010tf},  thereby providing estimates for the parameters $\hat M^2$, $\hat m^2$ and $\lambda^4$. Using these estimates, this propagator displays 2 complex conjugate poles, indicative of the gluon not being a physical degree of freedom. Moreover, the corresponding  K\"{a}ll\'{e}n-Lehmann (KL) representation for $D(p^2)$ displays a nonpositive spectral density, providing further evidence for the unphysical nature of the gluon \cite{Dudal:2013yva}. Nevertheless, despite the presence of unphysical poles, the confining gluon propagator \eqref{rgzgll} has been shown to provide good estimates of the masses of the first glueball states, through the evaluation of the correlation functions of the corresponding gauge-invariant composite operators \cite{Dudal:2010cd}.

It is worth underlining here that the Gribov parameter $\lambda$ induces a soft  breaking of the Becchi-Rouet-Stora-Tyutin (BRST) symmetry\footnote{We refer to Section 2 for some more details, next to the existing literature \cite{Dudal:2009xh}.}, i.e.~a breaking which is relevant only in the deep infrared region, not affecting the ultraviolet perturbative regime.  As it is well known, in a non-confining gauge theory the BRST invariance enables us to construct the physical subspace associated with the asymptotic fields, guaranteeing the unitarity of the $S$-matrix,    at least when working with a formal power series in the coupling.  On the other hand, in a confining theory there are no asymptotic states for the elementary fields.  In this sense, the existence of a soft breaking of the BRST symmetry is not  necessarily   incompatible with the confining character of the theory. Very recently, the first evidence for the existence of such a soft BRST breaking has been provided by lattice  numerical simulations  \cite{Cucchieri:2014via}.

The aim of the present work is that of using what we have learned in the gluon sector from the Gribov-Zwanziger framework to construct a non-perturbative renormalizable quark model accounting for both confinement and dynamical chiral symmetry breaking. As in the case of the gluon propagator, we shall be able to obtain a quark propagator which reproduces in an accurate way the lattice data. Besides a real pole, the resulting quark propagator exhibits complex conjugate ($cc$) poles.
It has been discussed in \cite{Alkofer:2003jj} that a quark propagator displaying exactly a set of $cc$ poles and a real pole does provide a good fit to the numerical solution of the quark propagator's Dyson-Schwinger equation. The appearance of $cc$ poles as an effective means to describe non-perturbative QCD dynamics has received anew attention during the past few years (cf. e.g. \cite{Fukushima:2012qa,benic,Benic:2013eqa,Fukushima:2013xsa,Su:2014rma,prep,Capri:2012hh,Baulieu:2009ha,Dudal:2013wja,Capri:2014bsa}).
Our goal here is not only to use a fit of the aforementioned type, but rather discuss a way to find a new vacuum that exactly displays this type of behaviour for the tree level quark propagator.
Furthermore, a detailed study of the obtained K\"{a}ll\'{e}n-Lehmann representation shows a violation of reflection positivity, a feature which hinders the standard physical description of asymptotic propagating quarks, being interpreted as an indication for quark confinement.

Similarly to the gluon sector, a soft breaking of the BRST symmetry shows up also in the quark matter sector, being related to the mass parameters appearing in the quark mass function of the model. Concerning now the dynamical chiral symmetry breaking, the model enables us to identify a set of local composite operators whose non-vanishing vacuum expectation values account for the chiral symmetry breaking, with the pion field corresponding to the associated (pseudo)Goldstone mode. Moreover, as in the case of the glueballs, the confining and positivity violating quark propagator allows for mass estimates of the meson spectrum, through the analytic evaluation of the spectral density for the correlation functions of local gauge invariant  mesonic operators, as illustrated in the case of the mass and decay constant of the $\rho$ meson.

In summary, we shall present a confining quark model exhibiting the following features:
\begin{itemize}
\item a non-perturbative quark propagator with a massive-like behaviour in agreement with lattice data;
\item  violation of reflection positivity, indicative of the unphysical nature of quarks;
\item  renormalizability, to describe quark dynamics over the whole momentum range, including its ultraviolet behaviour given by perturbative QCD;
\item  dynamical breaking of the chiral symmetry with the identification of the massless pion in the chiral limit;
\item bound state poles in adequate composite operators, allowing for estimates of meson masses.
\end{itemize}
The paper is organized as follows. In Section 2  we remind briefly the Gribov-Zwanziger framework and we  present the construction of the confining quark model. Section 3 is dedicated to the description of the dynamical chiral symmetry breaking and to the identification of the pion mode in our model. In Section 4 we give a detailed account of the violation of the reflection positivity of the quark propagator.  Section 5 outlines the analytic evaluation of the $\rho$ meson spectral function and the estimate of its mass and decay constant. Finally, a summary and outlook can be found in Section 6.   We have collected some technical details in a set of Appendices.

\section{A confining quark model}
In this Section, the model is constructed emphasizing the origin of the non-perturbative gauge and quark dynamics. The reduction to QCD in the ultraviolet regime and the construction of an effective action for the infrared region will be also discussed.

\subsection{Non-perturbative gluons via the Gribov horizon}
We initiate from the QCD action in $4d$ Euclidean space\footnote{We did not write bare quark masses, even though they will be taken into account later on in the paper. For  simplicity of presentation, we work with one flavour here, see the end Section 3 for a comment related to this. The method is however straightforwardly generalizable to the multiflavour case.}, but we add an extra piece:
\begin{equation}\label{e1}
    S=\int \d^4x \left(\frac{1}{4}F_{\mu\nu}^2\right)+S_{Landau}+\int \d^4x\left(\opsi\D  \psi -\underline{\overline\lambda^{ai}\p_\mu D_\mu^{ab}\lambda^{bi}-\overline\xi^{ai}\p_\mu D_\mu^{ab}\xi^{bi}-\p_\mu\xi^{ai}gf^{acb}D_\mu^{cm}c^m \lambda^{bi}}\right).
\end{equation}
The $\overline\lambda^{ia}$, $\lambda^{ia}$ are a set of (anticommuting) spinor fields while $\overline\xi^{ia}$, $\xi^{ia}$ a set of (commuting) spinor Grassmann fields. They carry a double color index $(i,a)$, with $i$, resp.~$a$ referring to the fundamental, resp.~adjoint representation. Here, $D_\mu^{ab}=\p_\mu \delta^{ab}-gf^{abc}A_\mu^c$ is the covariant derivative in the adjoint representation.

This particular setup was first introduced in \cite{Capri:2014bsa,Capri:2014fsa} in order to couple the nontrivial dynamics related to the Landau gauge Faddeev-Popov operator, $-\p_\mu D_\mu^{ab}$, to non-perturbative quark dynamics. To  describe infrared gluon dynamics (which is most probably a crucial factor for the dynamics behind chiral symmetry breaking), a non-perturbative Landau gauge fixing as the (Refined) Gribov-Zwanziger scheme can be selected \cite{Gribov:1977wm}, but also other schemes like that of \cite{Serreau:2012cg} are possible. For the RGZ case, we have in particular
\begin{eqnarray}\label{e2}
    S_{Landau}&=&\int \d^4x \left(b^a\p_\mu A_\mu^a+\overline c^a \p_\mu D_{\mu}^{ab}c^b+\overline\varphi_\mu^{ac}\p_\nu D_\nu^{ab}\varphi_\mu^{bc}-\overline\omega_\mu^{ac}\p_\nu D_\nu^{ab}\omega_\mu^{bc}-gf^{amb}\p_\nu\omega_\mu^{ac}D_\nu^{mp}c^p\varphi_\mu^{bc}\right)\nonumber\\
    &&\!\!\!+\int \d^4x\left(\gamma^2 gf^{abc}A_\mu^a(\varphi_\mu^{bc}+\overline\varphi_\mu^{bc})-\hat M^2(\overline\varphi_\mu^{ab}\varphi_\mu^{ab}-\overline\omega_\mu^{ab}\omega_\mu^{ab})+\frac{\hat m^2}{2}A_\mu^aA_\mu^a-4\gamma^4(N^2-1)-\frac{\zeta(g^2)}{2}\hat m^4\right)
\end{eqnarray}
where $\gamma^2$ is a mass parameter related to the restriction of the gauge field path integration to the Gribov region $\Omega$, while $\tilde M^2$ and $\tilde m^2$ are dynamically generated mass scales which stabilize the vacuum \cite{Gribov:1977wm}. The vacuum term $\propto m^4$ is necessary to ensure compatibility with a linear renormalization group equation \cite{Gribov:1977wm,Verschelde:2001ia}, with $\zeta(g^2)$ a calculable renormalization group function. In general, $\gamma^2$ must obey a gap equation, given by
\begin{equation}
\frac{\p E_{vac}}{\p \gamma^2}=0\, .
\label{gapeq}
\end{equation}
 In the absence of quarks, the RGZ dynamics is well compatible with lattice data\footnote{In the presence of dynamical quarks, the RGZ gluon fits have not yet been studied in detail.} \cite{Dudal:2010tf}. We recall here that the Landau gauge $\p_\mu A_\mu^a=0$ admits multiple solutions, thereby invalidating the standard Faddeev-Popov quantization. An infinitesimal gauge transform of $A_\mu^a\to A_\mu^a+D_\mu^{ab}\omega^b$ will respect the Landau gauge if $-\p_\mu D_\mu^{ab}\omega^{b}=0$, i.~e.~when the Faddeev-Popov operator $-\p_\mu D_\mu^{ab}$ has (normalizable) zero modes. The Gribov region $\Omega$ is defined via
\begin{equation}\label{grr}
  \Omega=\left\{A_\mu^a\left|\p_\mu A_\mu^a=0\,, \p_\mu D_\mu^{ab}>0\right.\right\}
\end{equation}
and the restriction to $\Omega$ thus removes the (at least the infinitesimal) gauge copies. The action  \eqref{e2} -- together with the gap equation \eqref{gapeq}  -- implements such restriction in a practical fashion \cite{Gribov:1977wm}.

The additional (fermion) fields appearing in \eqref{e1} are perturbatively trivial, as the underlined piece constitutes a unity. The first two terms can be integrated out exactly and cancel each other's contribution, while the third term $\p_\mu\xi^{ai}gf^{acb}D_\mu^{cm}c^m \lambda^{bi}$ does not play any role as it cannot be coupled to ghost number zero quantities because the ghost field $c$ appears without its antighost partner $\oc$. It must however   be   present to ensure   the   invariance of   the   new piece in the action under the (extended) BRST transformation
\begin{eqnarray}\label{brstn}
  sA_\mu^a&=&-D_\mu^{ab}c^b\,,\quad sc^a=\frac{g}{2}f^{abc}c^b c^c\,,\quad s\overline c^a=b^a\,,\quad sb^a=0\,,\nonumber\\
  s\overline\omega_\mu^{ab}&=&\overline\varphi_\mu^{ab}\,,\quad s \overline\varphi_\mu^{ab}=0\,,\quad s\varphi_\mu^{ab}=\omega_\mu^{ab}\,,\quad s\omega_\mu^{ab}=0\,,\nonumber\\
  s\psi^i&=& -igc^aT^a_{ij}\psi^j\,,\quad s\opsi^i= -ig \opsi^j c^a T^a_{ji}\,,\nonumber\\
    s\overline\xi^{ai}&=&\overline\lambda^{ai}\,,\quad s\overline\lambda^{ai}=0\,,\quad s\lambda^{ai}=\xi^{ai}\,,\quad s\xi^{ai}=0\,.
\end{eqnarray}
The action is thus perturbatively equivalent to the case as if the new fermion fields would be absent\footnote{In the recent work \cite{Kaplan:2013dca}, a similar premise was adopted to obtain ``extended QCD''. }, in particular should its symmetry content be. We treat the new fields as chiral singlets, as their quadratic form is not of the usual kind. We nevertheless recover the chiral symmetry under
\begin{equation}\label{e1bis}
    \delta_5 \psi= i\,\gamma_5\psi\,,\quad \delta_5\opsi=i\,\opsi\gamma_5\,,\quad \delta_5(\text{rest})=0.
\end{equation}

\subsection{  Soft BRST breaking and emergent non-perturbative quark dynamics}

We now point out that the restriction to the Gribov region softly\footnote{With softly, we mean proportional to a mass parameter. This also entails that at large momenta, such mass scale should become irrelevant, returning the physics back to the standard Faddeev-Popov one.} breaks BRST symmetry, since
\begin{equation}\label{bbre}
  s S_{Landau}=\int \d^4x\left(\gamma^2gf^{abc}D_\mu^{ak}c^k(\varphi_\mu^{bc}+\overline\varphi_\mu^{bc})+\gamma^2gf^{abc}A_\mu^a\omega_\mu^bc\right)
\end{equation}
This BRST breaking has recently received support from lattice simulations \cite{Cucchieri:2014via}, in which case the restriction to the Gribov region is always present. Indeed, the Landau gauge is imposed numerically via a minimization, along the gauge orbit, of $\int \d^4x~  A_\mu^2$, unequivocally implying that $\p_\mu A_\mu^a=0$ (vanishing first order variation) and $-\p_\mu D_\mu^{ab}>0$ (positive second order variation).

Since the gluon sector, with its BRST breaking, is coupled to the quarks and to the new fermion sector via the Faddeev-Popov operator, it can be expected that the BRST breaking effects will also manifest themselves in the whole fermion sector. In the current paper, we will therefore adopt this as our working hypothesis. In fact, we will see that this also offers an opportunity to discuss chiral symmetry breaking and confinement (in the sense of violation of reflection positivity by the fundamental degrees of freedom and existence of physical bound states), providing us with a quark propagator that can describe quite well the corresponding lattice data (see also \cite{Capri:2014bsa}). The soft breaking of the BRST invariance in the gluon sector can be reinterpreted by means of an exact non-perturbative
Ward identity which leads to a non-perturbative modification of the standard BRST transformations, see \cite{Capri:2015ixa,Capri:2015pfa}. This non-perturbative BRST \emph{unbroken symmetry} can in principle be generalized to the quark sector, thereby providing a BRST invariant extension of the here described non-perturbative quark dynamics. We leave a further study of these matters to future work.

In our approach, the transmission of soft BRST breaking to the quark sector is encoded in the generation of nonzero condensates of composite operators involving the matter and the auxiliary fields.
 Consider next the local composite operators (it is useful to record the mass dimensions: $\dim[\psi,\opsi]=3/2$ and $\dim [\lambda,\overline\lambda,\xi,\overline\xi]=1$)
\begin{eqnarray}\label{lcos}
  \mathcal{O}_1&=&\overline\lambda^{ai} T^a_{ij}\psi^j + \overline\psi^i T^a_{ij}\lambda^j\,,\nonumber\\
  \mathcal{O}_2&=&\overline\lambda^{ai}\lambda^{ai}+\overline\xi^{ai}\xi^{ai}
\end{eqnarray}
with $\dim[\mathcal{O}_1]=5/2$ and $\dim[\mathcal{O}_2]=2$. The mixed fermion condensate $\braket{\mathcal{O}_1}$ serves as an order parameter for chiral symmetry, since we have
\begin{eqnarray}\label{ord}
\mathcal{O}_1=-\delta_5\pi\quad\text{with}\quad\pi=-i~\left(\overline\lambda^{ai}T^a_{ij}\gamma_5\psi^j+\opsi^iT^a_{ij}\gamma_5\lambda^j\right)
\end{eqnarray}
We shall later on show that $\pi$ corresponds to the pion, viz.~a massless pseudoscalar in the chiral limit.

Let us now discuss in brief how to construct the underlying effective action, $\Gamma$, for the non-perturbative dynamics related to $\braket{\mathcal{O}_1}$ and $\braket{\mathcal{O}_2}$. We introduce 2 scalar sources, $J$ and $j$, coupled to $\mathcal{O}_1$ and $\mathcal{O}_2$, with $\dim[J]=3/2$, $\dim[j]=2$,
\begin{equation}\label{act10}
    S\to S+ \int \d^4x\left(J\mathcal{O}_1+j\mathcal{O}_2-\chi(g^2)\frac{j^2}{2}\right)
\end{equation}
as derivation w.r.t.~the sources allows to define the quantum operators. As noted before, the function $\chi(g^2)$ is indispensable for a homogeneous linear renormalization group for $\Gamma$, while its value can be consistently determined order by order, making it a function of the coupling $g^2$. It reflects the vacuum energy divergence $\propto j^2$. We refer to \cite{Verschelde:1995jj,Knecht:2001cc} for the seminal papers plus toolbox concerning this so-called Local Composite Operator (LCO) method.

As an important asset of $4d$ quantum field theory is its multiplicative renormalizability, this property needs to be established for eq.~\eqref{act10}. Using a more general set of sources, this can be proven to all orders of perturbation theory, see for instance \cite{Capri:2014fsa}. In particular, an important consequence of the proof is the absence of pure vacuum terms in $J$. Although RG controllable, we need a workable action description as the nonlinear $\chi(g^2)\frac{j^2}{2}$ clouds the energy interpretation and outruns standard 1PI formalism \cite{Banks:1975zw}. This can be overcome by a double Hubbard-Stratonovich (HS) unity:
\begin{equation}\label{hs1}
1=\mathcal{N}\int\left[\d\sigma d\Sigma\right] e^{-\frac{1}{2\chi}\int \d^4x\left(\sigma-\chi j+\mathcal{O}_2\right)^2}e^{-\frac{1}{2\Lambda}\int \d^4x\left(\Sigma-\Lambda J+\mathcal{O}_1\right)^2}\,,
\end{equation}
leading to an equivalent action,
\begin{eqnarray}\label{hs2}
    S_f+S_{j,J}&\equiv&\int \d^4x \left(\frac{1}{4}F_{\mu\nu}^2+\opsi\D  \psi -\overline\lambda^{ai}\p_\mu D_\mu^{ab}\lambda^{bi}-\overline\xi^{ai}\p_\mu D_\mu^{ab}\xi^{bi}-\p_\mu\xi^{ai}gf^{acb}D_\mu^{cm}c^m \lambda^{bi}\right)\nonumber\\
    &&\hspace{-1cm}+\int \d^4x\left( \frac{\sigma^2}{2\chi}+\frac{\sigma}{\chi}\mathcal{O}_2+\frac{\mathcal{O}_2^2}{2\chi}+\frac{\Sigma^2}{2\Lambda}+\frac{\mathcal{O}_1^2}{2\Lambda}+\frac{\Sigma}{\Lambda}\mathcal{O}_1-\sigma j-\Sigma J+\Lambda J^2\right)\,.
\end{eqnarray}
Amusingly, contrary to its usual purpose, the HS transformation consistently introduces quartic fermion interactions that are harmless from the renormalization viewpoint, this in contrast to the NJL quartic fermion interactions.  We may discard the $J^2$ term in \eqref{hs2}, since it is irrelevant for RG purposes: without changing the physics, we could have added a canceling $-\Lambda J^2$ to the action \eqref{act10}. The sources now appear linearly. Acting with
\begin{equation}
\left.\frac{\p}{\p \{j,J\}}\right|_{j=J=0}
\end{equation}
on both partition functions, i.e.~before and after the HS transformation, provides with the correspondences:
\begin{equation}\label{ord2}
 \braket{\Sigma}=-\braket{\mathcal{O}_1} \quad {\rm and}\quad \braket{\sigma}=-\braket{\mathcal{O}_2}.
 \end{equation}
 $\Lambda$ is a mass dimension 1 parameter, necessary to end up with the appropriate mass dimensions throughout, since $\dim[\sigma]=2$, $\dim[\Sigma]=5/2$. $\Lambda$ will not enter any physical result if we were to compute exactly, as the underlying transformation \eqref{hs1} constitutes just a unity.  In a loop expansion, $\Lambda$ will unavoidably enter any calculated quantity. However, assuming that passing to higher order one gets closer to the exact result, which encompasses $\Lambda$-independence, we can fix $\Lambda$ in a case-by-case scenario by the principle of minimal sensitivity (PMS) \cite{Stevenson:1981vj}: we look for solutions of $\frac{\p E_{vac}}{\p \Lambda}=0$ or higher derivatives if the latter eq.~has no zeros. In principle, $\Gamma$ itself could be examined using the background field method \cite{Jackiw:1974cv}; then we do no longer need the sources and can set them to zero.

Finally, if the dynamics would decide that $\braket{\Sigma}=\braket{\sigma}=0$, then we are dealing with nothing else than QCD without D$\chi$SB, as the trivial and thus irrelevant unities can be integrated out. If, on the contrary, $\braket{\Sigma}\neq0$ by means of dimensional transmutation, we find ourselves in a vacuum where chiral invariance is dynamically broken. We draw attention to the invariance of the action itself, $\delta_5S_f=0$, since naturally $\delta_5\Sigma=-\delta_5\mathcal{O}_1$. The r\^{o}le of $\braket{\sigma}$ is to furnish a dynamical mass for the auxiliary fermion fields.

It is immediate to see that dynamically obtaining a mass scale coupled to the operator  $\mathcal{O}_1$ will correspond to a (soft) BRST breaking in the fermion sector, as
    \begin{equation}\label{bbr3}
      s\mathcal{O}_1=ig^2\overline\lambda^{ai}T^a_{ij}c^bT^b_{jk}\psi^k-ig^2\opsi^k c^b T^b_{ki}T^a_{ij}\lambda^{aj}-g\opsi^i T^a_{ij}\xi^{aj}
    \end{equation}
Given the soft BRST breaking term in the gluon sector (due to the Gribov restriction), this provides a scenario of ``mediated'' soft BRST breaking into the fermion sector, thereby giving a dynamical ground to the study presented in \cite{Capri:2014bsa}. We have in this way a dynamical picture for quark confinement via positivity violation as a result of the non-perturbative gauge dynamics.

With a bare quark mass $\mu$, the tree level quark propagator of the model yields
\begin{equation}\label{qp}
   \Braket{\psi \opsi }_p=\frac{i\pp+\mathcal{M}(p^2)}{p^2+\mathcal{M}^2(p^2)}\,,\quad
    \mathcal{M}(p^2)=\frac{\braket{\Sigma}^2/\Lambda^2}{p^2+\braket{\sigma}}+\mu
\end{equation}
The momentum dependent non-local  mass function $\mathcal{M}(p^2)$ is a result of the chiral symmetry breaking, in the sense that $\mathcal{M}(p^2)=0$ if chiral symmetry would be realized in the vacuum, since then $\braket{\Sigma}=-\braket{\mathcal{O}_1}=0$ as $\mathcal{O}_1$ is the chiral variation of another operator, eq.~\eqref{ord}. The foregoing tree level functional form has been applied in fits to lattice quark propagators in \cite{Furui:2006ks,Parappilly:2005ei,Burgio:2012ph,Rojas:2013tza}.
Using the data of \cite{Parappilly:2005ei} (in particular, their Figure~5), we consider light quarks, with current mass $\mu=0.014~\text{GeV}$. The dynamical quark mass can be fitted excellently with
\begin{equation}\label{fit}
\mathcal{M}(p^2)=\frac{M^3}{p^2+m^2}+\mu~\text{with}~M^3=0.1960(84)~\text{GeV}^3\,, m^2=0.639(46)~\text{GeV}^2 \quad (\chi^2/\text{d.o.f.}~=~1.18)\,.
\end{equation}
see also Figure~1.
\begin{figure}[t]
   \centering
       \includegraphics[width=9cm]{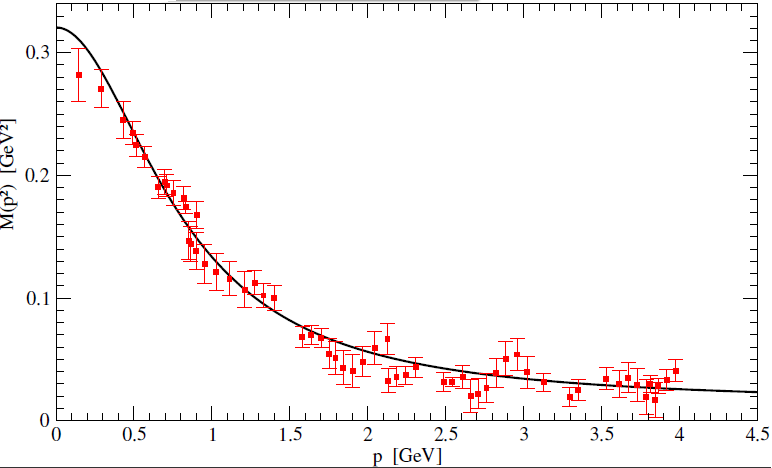}
               \caption{Lattice quark mass function \cite{Parappilly:2005ei} with its fit $\mathcal{M}(p^2)$.}\label{figfit}
\end{figure}

This corroborates the relevance of our model, since we end up with a quark propagator in a nontrivial vacuum, whose functional form is consistent with the non-perturbative lattice counterpart. Moreover, we did not make any sacrifices with respect to the renormalizability and, in the absence of condensation, our action is equivalent to that of perturbative QCD.  Moreover, as we shall see below, our quark propagator explicitly displays a positivity violation, thereby showing that the quark degrees of freedom cannot be part of a physical subspace; this can be interpreted as a signal of confinement. Furthermore, the physical spectrum of the model will be composed of bound states of these unphysical fundamental fields, as we will discuss in Section \ref{meson}.

The same tree level quark propagator was already discussed before in \cite{Baulieu:2009xr} albeit using an {\it ad-hoc} renormalizable model.
Before continuing, we wish to point out here that a momentum dependent quark mass also showed up in an instanton based analysis of the QCD vacuum \cite{Diakonov:1985eg,Diakonov:1987ty} or by solving the Dyson-Schwinger equation for the quark propagator \cite{Bhagwat:2002tx,Bhagwat:2003vw}.

For now, we will take advantage of the foregoing lattice studies to fix our condensates. This will provide us with a tree level quark propagator, with the global form factor $\mathcal{Z}\equiv1$. Including loop corrections on top of the non-perturbative vacuum will lead to $\mathcal{Z}(p^2)$, as also seen in e.g.~Figure~5 of \cite{Parappilly:2005ei}, which however only deviates mildly from $1$ over a large range of momenta: the tree approximation $\mathcal{Z}=1$ appears to be valid. In the future, we hope to come back to an explicit self-consistent computation and analysis of the effective action. As the LCO method necessitates the computation of $(n+1)$-loop divergences/anomalous dimensions to get the correct (renormalization group compatible) $n$-loop result, see \cite{Verschelde:1995jj,Knecht:2001cc}, this is a tremendous effort that deserves separate attention.  For the benefit of the reader, let us give a short overview here of the LCO method. We ought to depart from the action $S_f$, eq.~\eqref{hs2}, and we will work in the background field formalism. We can set the sources to zero and write
\begin{equation}\label{bfm1}
  \sigma=\braket{\sigma}+\tilde\sigma\,,\quad \Sigma=\braket{\Sigma}+\tilde\Sigma\,,
\end{equation}
where the tilde-fields are the fluctuations around a possible nonzero vacuum expectation value. We can then integrate out all fluctuations to arrive at the vacuum effective action,
\begin{equation}\label{bfm2}
  \Gamma(\braket{\sigma},\braket{\Sigma},\Lambda,\chi)\,.
\end{equation}
$\Gamma$ will depend on the condensates $\braket{\sigma}$, $\braket{\Sigma}$, as usual, but also on the parameters introduced via the double HS unity in \eqref{hs1}, namely: the hitherto free parameter $\chi$ and the (renormalization-group invariant) PMS scale $\Lambda$. For any value of $\Lambda$, the functional $\Gamma$ should obey a standard linear homogeneous renormalization group equation,
\begin{equation}\label{bfm3}
  \left[\omu\frac{\p}{\p \omu}+\beta(g^2)\frac{\p}{\p g^2}+\gamma_\sigma(g^2)\frac{\p}{\p \braket{\sigma}}+\gamma_\Sigma(g^2)\frac{\p}{\p \braket{\Sigma}}\right]\Gamma=0\,,
\end{equation}
where $\omu$ is the renormalization scale, and $\beta(g^2)$, $\gamma_\sigma(g^2)$ and $\gamma_\Sigma(g^2)$ the anomalous dimensions of $g^2$, $\sigma$ and $\Sigma$. As explained in \cite{Verschelde:1995jj,Knecht:2001cc}, this is not automatically so, unless $\chi$, that also runs with the scale $\omu$, is fixed order by order in perturbation theory in a unique (analytical) way via a Laurent series expansion in the coupling,
\begin{equation}\label{bfm4}
  \chi=\frac{\chi_0}{g^2}+\chi_1+\chi_2 g^2+\ldots\,.
\end{equation}
In practice, the coefficients $\chi_i$ can be obtained from imposing the renormalization-group equation \eqref{bfm3} to hold order by order. As such, the $\chi_i$ get related -- via the other anomalous dimensions -- to the $\omu$-dependent logarithms (and thus to the renormalized divergences) present in the vacuum functional. The price one has to pay for this, as said before, is that e.~g.~$\chi_0$ that enters the tree level vacuum energy is related to a one-loop logarithm term, etc.~This will always require $(n+1)$-loop computations to fix the $\chi_i$ up to $n$-loops.

Notice that, unlike the a priori free mass scale $\Lambda$, the dimensionless parameter $\chi$ cannot be chosen at will. It must obey its renormalization group equation. $\Lambda$ on the other hand will not run.

\section{Dynamical $\chi$SB and massless pion in the chiral limit of the confining model}
Let us now scrutinize whether we can introduce a pion field. We consider $\braket{\Sigma}\neq0$ and the already introduced field $\pi$ [we assume the chiral limit here, $\mu=0$]. Using $S_f$, the chiral current is easily derived to be $j_\mu^5=\opsi \gamma_\mu\gamma^5\psi$. In what follows, we shall adapt the standard derivation, presented in e.g.~\cite{Pokorski:1987ed}.
The correlation function
    \begin{equation}
        \mathcal{G}_\mu(x-y)=\braket{j_\mu^5(x)\pi(y)}
    \end{equation}
    is subject to
    \begin{equation}
    \p_\mu\mathcal{G}_\mu(x-y)=\delta(x-y)\braket{\Sigma}
    \end{equation}
    using either a path integral or a current algebra argument. Fourier transforming and using the Euclidean invariance yields
    \begin{equation}\label{conn}
    \mathcal{G}_\mu=\frac{p_\mu}{p^2}\braket{\Sigma}
    \end{equation}
    To close the argument, we consider the $\mathcal{S}$-matrix element of the current destroying a pion state,
    \begin{equation}
    \braket{j_5^\mu(x)\pi(p)}\propto p_\mu e^{-ipx}
     \end{equation}
     which is related to the amputated propagator when the pion is put on-shell. Assuming a pion mass $m_\pi$,  we would get
     \begin{equation}
     \braket{j_5^\mu(x)\pi(p)}\propto\lim_{p^2\to -m_\pi^2} (p^2+m_\pi^2) \mathcal{G}_\mu(p) e^{-ipx}
      \end{equation}
      with pion propagator
      \begin{equation}
      \braket{\pi\pi}_{p^2\sim -m_\pi^2}\propto\frac{1}{p^2+m_\pi^2}
      \end{equation}
      Recombination provides us with
      \begin{equation}
      \lim_{p^2\to -m_\pi^2} (p^2+m_\pi^2) \mathcal{G}_\mu(p)\propto p_\mu
      \end{equation}
      and having already shown eq.~\eqref{conn}, we must require $m_\pi^2=0$, i.e.~$\pi$ does describe a massless particle.\newline It is instructive to notice that, using the tree level action stemming from \eqref{hs2} in the condensed phase, we can rewrite
      \begin{equation}
      \pi_{lowest\,order}\propto\opsi\frac{\Sigma/\Lambda}{-\p^2+\braket{\sigma}}\gamma^5\psi
      \label{nonlocalpi}
       \end{equation}
       by means of the equations of motion of the auxiliary fermions. We recognize a nonlocal version of the usual pseudoscalar pion field. This is not a surprise, since the action \eqref{hs2} itself becomes a nonlocal quark action upon integrating out the extra fermion fields in the condensed phase.

At this point, one may wonder what is the relation between the condensate $\braket{\Sigma}$ introduced here and the standard chiral condensate $\braket{\bar\psi\psi}$.
As already discussed above, when $\braket{\Sigma}=0$,  the theory is chirally symmetric and the auxiliary fermion fields can be integrated out trivially, yielding the standard QCD action with massless quarks. It is then straightforward to see that the chiral condensate $\langle\bar\psi\psi \rangle$ must be zero order per order in perturbation theory whenever $\braket{\Sigma}$ vanishes. We remind here the goal of our model is to work with perturbation theory on top of a non-perturbative vacuum. If the vacuum is unchanged compared to standard perturbative QCD, we will of course not have chiral symmetry breaking in perturbation theory. The opposite is also true: if $\braket{\Sigma}\ne 0$, then the chiral invariance of the (now non-perturbative) vacuum is lost and thus the chiral condensate $\braket{\bar\psi\psi}$ will be nonzero order per order. For example, in the chiral limit ($\mu=0$) and working at leading (quadratic) order, we find, analogously to \cite{Capri:2014xea},
\begin{equation}
\braket{\opsi\psi}=\int \frac{\d^4p}{(2\pi)^4} \text{Tr}\braket{\opsi(-p)\psi(p)}=\int \frac{\d^4p}{(2\pi)^4}\frac{\mathcal{M}(p^2)}{p^2+\mathcal{M}^2(p^2)}\propto \braket{\Sigma}\,.
\end{equation}
Therefore, one concludes that  $\braket{\Sigma}$  and $\braket{\bar\psi\psi}$ are both order parameters for the chiral transition, and as such they are equally well-fitted to describe D$\chi$SB.

Strictly speaking, the above analysis is not entirely correct as we kept using the single flavour setup, in which case it is known that the chiral anomaly hampers the chiral symmetry in the single quark flavour sector. The purpose of this section was however to merely to illustrate that the new pion field is massless based on a current algebra argument. A completely analogous argument can be written down for e.g.~the physically more relevant two flavour sector, in which case the chiral anomaly does not interfere with the masslessness of the usual $\pi^{0,\pm}$ pion fields.

It is anyhow interesting to point out that the anomaly itself is unchanged in our setup. Indeed, unlike the NJL model, we keep the gauge fields present. There is thus no need to model in anomaly driven (chirally non-invariant) interactions via e.g.~a 't Hooft determinant or 6-quark vertices. The chiral transformation itself is also unchanged, cf.~the natural extension written down in eq.~\eqref{e1bis}. A first consequence is the already mentioned fact that the standard expressions for the chiral current will not be affected compared to the QCD case. As a second consequence, we can follow the classic Fujikawa path integral derivation \cite{Fujikawa:1979ay} to find that the potential anomalous contribution to the currents' divergence will be proportional to the appropriate pseudoscalar quantity $FF^\ast$. Said otherwise, nothing will change in our model in the anomalous sector compared to QCD itself.

\section{Violation of reflection positivity for quarks}
Having accomplished a consistent non-perturbative quark model with the correct chiral behavior, it remains to address the issue of reflection positivity and its violation.
The particular goal of the current Section is that of showing that the quark model proposed indeed presents quark degrees of freedom which violate this property being thus absent from the physical spectrum. Nevertheless, for the benefit of the reader, we shall briefly discuss first on general grounds the property of reflection positivity for Dirac fermions\footnote{A previous account of positivity violation in the case of quarks can be found in the context of Dyson-Schwinger equations \cite{Alkofer:2008tt}.} and then present the results for our model.

Notice however we do not proclaim that positivity violation of quarks is equivalent to confinement, the latter being anyhow a delicate topic to define in presence of dynamical quarks \cite{Greensite:2011zz}. We are not able to derive, via the Wilson loop, a linear piece in the interquark potential for sufficiently small interquark distance (to avoid pair production) using a fixed order of perturbation theory with our model. To our knowledge, there is actually no functional method capable so far of providing such potential, unless strongly infrared divergent propagators/vertices (a fact not supported by lattice data)  are used as in \cite{Alkofer:2006gz}. Nevertheless, non-perturbative propagators can be connected to confinement via the Polyakov loop at finite temperature, see \cite{Fukushima:2012qa,Canfora:2015yia} for the example with a gluon propagator of the Gribov type. Our main observation is that, in the absence of quark positivity, quarks cannot be observable quantities. A similar viewpoint was scrutinized in more detail in \cite{Krein:1990sf,Roberts:2007ji}.

In quantum field theory a given field may be associated with a physical particle (with a propagating asymptotic state) if one is able to define a probability of particle propagation through 2-point correlation functions of this field. In general, necessary properties of a probabilistic description are well known: the conditions of positivity and unitarity.
Let us focus on the positivity condition. A field with physical particle interpretation must thus display a 2-point correlation function that can be associated with a positive norm.
If that is not the case for any given field, than the associated quantum field theory should be regarded as a statistical description of these objects, which of course cannot receive a particle interpretation and as such shall not be part of the physical spectrum predicted by the theory.

It is an intricate problem to define these conditions in formal terms, but the positivity condition may be formalized as the Osterwalder-Schrader  axiom of reflection positivity \cite{Osterwalder:1973dx,Osterwalder:1974tc}:
\begin{equation}
\int \d^4x\d^4y \bar f (-x_0,\vec{x})\Delta(x-y)f(y_0,\vec{y}) \geq 0
\end{equation}
where $\Delta(x-y)$ is (a scalar function extracted from) the two-point correlation function in coordinate space with explicit translation invariance and $f(x_0,\vec{x})$ represents any complex-valued test function with support for positive times. In the mixed coordinate-momentum representation we arrive at:
\begin{equation}
\int_0^{\infty} \d t\d t' \bar f (t',\vec{p})\Delta(-(t-t'),\vec{p})f(t,\vec{p}) \geq 0
\,,
\end{equation}
which should be satisfied for all values of three-momentum $\vec{p}$. It is clear that if $\Delta(-(t-t'),\vec{p})$ is negative within any domain, than it is easy to find a test function that will pinpoint this negative region and violate the inequality. The demonstration of posivity violation may be done thus by looking for a negative region of $\Delta(-(t-t'),\vec{p})$. For zero momentum, one has in particular the condition of positivity of the Schwinger function:
\begin{equation}
\Delta(t)=
\int \d p_4  e^{itp_4}  \Delta(p^2=p_4^2)\,,
\label{posSF}
\end{equation}
where we have used the fact that $\Delta$ is a scalar function extracted from the corresponding propagator,   which is   in general a function of $p^2$.

The nontrivial task that remains is to find out the appropriate scalar part of the propagator. For non-scalar fields this is not as straightforward. Our strategy here (following \cite{ColemanNotes}) is to write down the spectral representation for standard Dirac fermions, keeping track of the positive definite quantities in the process. In the end,
one identifies   two   scalar Schwinger functions $\Delta_+(t)$ and $\Delta_-(t)$ that   must   satisfy positivity conditions and   relates   them to the vector and scalar Schwinger functions $\Delta_v(t)$ and $\Delta_s(t)$, the  definition of which follow shortly. The scrutinization of the standard case provides then the positivity conditions to be tested for in our non-perturbative quark propagator in order to prove the absence of asymptotic particle description and the existence of confinement in this sense.

We may summarize the procedure of search for positivity violation in a fermionic propagator in Euclidean space in the following way (a detailed derivation can be found in Appendix \ref{ApPos}). The input is the propagator in momentum space:
\begin{eqnarray}
\Delta_E(\slashed{p})&=&-i\slashed{p}\sigma_v(p^2)+\sigma_s(p^2)\,.
\end{eqnarray}
The Schwinger functions are then computed using simple Fourier transforms:
\begin{eqnarray}
\Delta_v(t)&=&\frac{1}{2\pi}\int_{-\infty}^{\infty}\d p {\rm e}^{ipt}\sigma_v(p^2),
\\
\Delta_s(t)&=&\frac{1}{2\pi}\int_{-\infty}^{\infty}\d p {\rm e}^{ipt}\sigma_s(p^2),
\end{eqnarray}
which in turn should be tested under the two positivity conditions (valid for all times $t$):
\begin{enumerate}
\item $\Delta_v(t) \geq 0$,
\item $-\partial_t \Delta_v(t) \geq \Delta_s(t)$.
\end{enumerate}

\subsection{Results for the Schwinger functions of the non-perturbative fermion propagator and positivity violation}
From the quark propagator in our confining model, eq.\eqref{qp}, the appropriate scalar functions to be used to test for positivity may be read out:
\begin{eqnarray}
\sigma_v(k^2)&=&
\frac{1}{k^2+\mathcal{M}^2(k^2)}\,,
\\
\sigma_s(k^2)&=&
\frac{\mathcal{M}(k^2)
}{k^2+\mathcal{M}^2(k^2)}\,,
\end{eqnarray}
from which one can easily compute the Schwinger functions $\Delta_I(t)$ through Fourier transforms.

\begin{figure}[h!]
   \centering
       \includegraphics[width=9cm]{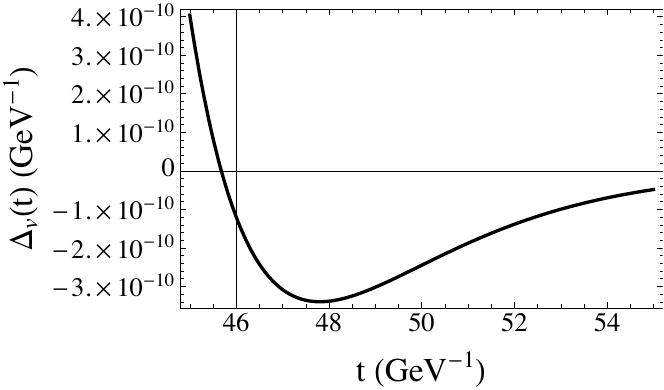}
               \caption{First positivity condition for the quark propagator in the confined model.}\label{Deltav1}
\end{figure}

\begin{figure}[h!]
   \centering
       \includegraphics[width=9cm]{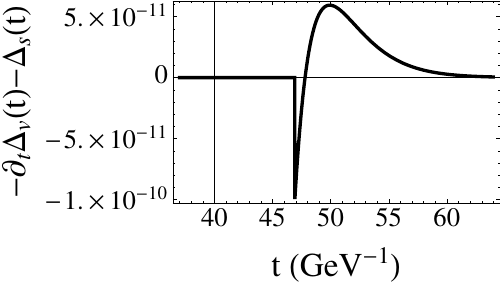}
               \caption{Second positivity condition for the quark propagator in the confined model.\label{Deltas1}}
\end{figure}

In Figures \ref{Deltav1} and \ref{Deltas1} we plot quantities that should satisfy the positive conditions derived above. One observes thus a clear violation of both positivity conditions by the tree-level quark propagator in our non-perturbative model, guaranteeing their absence from the physical spectrum of the theory, in agreement with what one expects from the confinement phenomenon.


Although the scale $t^*$ at which the quark positivity violation sets in appears to happen at a relatively large value of $t$, in units of inverse GeV, we are actually unaware
of any restriction applicable to the value of this scale. As we reminded earlier in this Section, the existence of a positivity violation of a given Euclidean correlation function forbids its physical particle interpretation in Minkowski space irrespective of the particular scale at which it has been detected.  Looking in more detail at the positivity requirement, this is actually defined through the use of test functions. Depending on the choice of the test function one uses, the zero-crossing might happen at quite different scales.
For the gluon, the available  positivity-violation studies (in lattice simulations or in continuum approaches like Dyson-Schwinger or Refined Gribov-Zwanziger) find a crossing scale of the same order of the QCD scale. This fact could induce to the expectation that both scales should be of the same order, but no direct link between this scale and the fundamental QCD scale is known.
Indeed, the only detailed work using a different approach to study the positivity violation in the quark sector is, to our knowledge, Ref. \cite{Alkofer:2003jj}, which combines
Dyson-Schwinger equations to fits of available lattice data at the time. Also there, the crossing seems to happen at relatively large values of $t$.
In the current analysis we also have systematic effects that are expected to shift the value of the scale $t^*$. Since we ignored the effects of $\mathcal{Z}(p^2)\neq 1$ (see the end of Section 2), including finite wave function renormalization effects can shift the value of $t^*$.
It is also worth recalling that we are referring, in our propagator fits, to lattice data related to larger-than-physical current quark masses. There is a factor 3--5 in the current quark mass, so this probably gives (using a scaling like the Gell Mann-Oakes-Renner relation) a factor of $\sqrt 3$--$\sqrt 5$ in the relevant hadronic scales and in the energy scale of the onset of quark confinement. A similar factor might influence $t^*$, despite the fact that $t^*$ is far from being a physical hadronic scale.

\section{Bound state spectrum in the confining quark model: the example of the $\rho$-meson\label{meson}}
As the proposed framework displays the desired chiral properties as well as the absence of asymptotic quarks, we should address further its physical spectrum, given by bound states of the unphysical fundamental degrees of freedom, i.e. the confined quarks.
As a representative example, let us construct and solve a gap equation for the (charged) $\rho^\pm$ meson mass and decay constant under suitable simplifying approximations. We consider degenerate up ($u$) and down ($d$) quarks with current mass $\mu=0.014~\text{GeV}$.
Again, our input will be the dynamical quark mass with parameters fitted from the lattice data (cf.~Figure \ref{figfit} and eq.~\eqref{fit}). We are ultimately interested in obtaining a pole in the charged $\rho$-meson channel, corresponding to a bound state.

\subsection{The charged $\rho$-meson correlator}
We may generalize the technology set out in \cite{Capri:2012hh} to the QCD case. The basic idea is to compute 2-point functions of appropriate composite operators and show that they can be associated with the propagation of a physical particle, i.e. that one can construct a K\"{a}ll\'{e}n-Lehmann spectral representation for these correlators.

We need the operators $\rho_\mu^-=\overline u \gamma_\mu d$, $\rho_\mu^+=\overline d \gamma_\mu u$ and their correlation function. We consequently consider
\begin{eqnarray}
\Braket{\rho_{\mu}^-(x)
\rho_{\nu}^+(y)}
&=&
\int\frac{\d^4k\d^4q\d^4k'\d^4q'}{(2\pi)^{16}}
\,
{\rm e}^{-i(k+q)\cdot x-i(k'+q')\cdot y}
\Braket{
\overline{u}(k)\gamma_{\mu}
d(q)\overline{d}(k')\gamma_{\nu}
u(q')}
\,,
\end{eqnarray}
where up to tree level we find
\begin{eqnarray}
\Braket{
\overline{u}(k)\gamma_{\mu}
d(q)\overline{d}(k')\gamma_{\nu}
u(q')}
&=&
{\rm Tr}\Big[
\delta(k+q') \Braket{u\overline u}_k
\gamma_{\mu}
\delta(k'+q)  \Braket{d\overline d}_q\gamma_{\nu}
\Big]
\,,
\end{eqnarray}
using the quark propagators in momentum space given in eq.~\eqref{qp}. Thence,
\begin{eqnarray*}
\Braket{
\rho_{\mu}^-(x)
\rho_{\nu}^+(y)}
&=&
\int\frac{\d^4k\d^4q}{(2\pi)^8}
\,
{\rm e}^{-i(k+q)\cdot(x-y)}
\frac{{\rm Tr}\Big\{
[i\slashed{k}+ \mathcal{M}_u(k^2)]
\gamma_{\mu}
[i\slashed{q}+ \mathcal{M}_d(q^2)]
\gamma_{\nu}
\Big\}
}{[k^2+ \mathcal{M}^2_u(k^2)][q^2+ \mathcal{M}^2_d(q^2)]}~=~
\int\frac{\d^4k}{(2\pi)^4}
\,{\rm e}^{-ik\cdot(x-y)}
\Braket{
\rho_{\mu}^-
\rho_{\nu}^+}_k
\,,
\end{eqnarray*}
with
\begin{eqnarray}
\Braket{
\rho_{\mu}^-
\rho_{\nu}^+}_k
&=&
\int\frac{\d^4q}{(2\pi)^4}
\,
\frac{{\rm Tr}\Big\{
[i\gamma_{\rho}(k_{\rho}-q_{\rho})+ \mathcal{M}_u\big((k-q)^2\big)]
\gamma_{\mu}
[i\gamma_{\sigma}q_{\sigma}+ \mathcal{M}_d(q^2)]
\gamma_{\nu}
\Big\}
}{[(k-q)^2+ \mathcal{M}^2_u\big((k-q)^2\big)][q^2+ \mathcal{M}^2_d(q^2)]}
\,.
\end{eqnarray}
Using standard results for the trace over $\gamma$-matrices,
\begin{eqnarray}
{\rm Tr}\Big[
\gamma_{\mu}\gamma_{\nu}
\Big]&=&
4\delta_{\mu\nu}\,,
{\rm Tr}\Big[
\gamma_{\rho}\gamma_{\mu}\gamma_{\sigma}\gamma_{\nu}
\Big]=
4[\delta_{\rho\mu}\delta_{\sigma\nu}
-\delta_{\rho\sigma}\delta_{\mu\nu}
+\delta_{\rho\nu}\delta_{\mu\sigma}
]\,,
{\rm Tr}\Big[
{\rm odd~number~of~\gamma's}
\Big]
=0
\,,
\end{eqnarray}
we arrive at
\begin{eqnarray}
\Braket{
\rho_{\mu}^-
\rho_{\nu}^+}_k
&=&
4 \int\frac{\d^4q}{(2\pi)^4}
\,
\frac{
-k_{\mu}q_{\nu}
-k_{\nu}q_{\mu}
+2q_{\mu}q_{\nu}
+\delta_{\mu\nu}(k\cdot q-q^2)
+\delta_{\mu\nu}\, \mathcal{M}_u\big((k-q)^2\big)\,
 \mathcal{M}_d(q^2)
}{[(k-q)^2+ \mathcal{M}^2_u\big((k-q)^2\big)][q^2+ \mathcal{M}^2_d(q^2)]}
\,.
\end{eqnarray}
In the case of degenerate up and down quark mass, the following correlator is transverse thanks to the EOMs and therefore guaranteed to describe a massive spin $1$ particle:
\begin{equation}\label{corr1}
  \braket{\rho_\mu^-\rho_\nu^+}_k=\frac{1}{3}\left(\delta_{\mu\nu}-\frac{k_\mu k_\nu}{k^2}\right)\braket{\rho_\rho^-\rho_\rho^+}_k\,.
\end{equation}
Explicitly, we have
\begin{eqnarray}
\Braket{
\rho_{\rho}^-
\rho_{\rho}^+}_k
&=&
8 \int\frac{\d^4q}{(2\pi)^4}
\,
f_{\rho}^{-+}(k,q)\,
\frac{1}{\Big[(k-q)^2\big[(k-q)^2+m^2\big]^2+\big\{M^3+\mu
\big[(k-q)^2+m^2\big]\big\}^2\Big]}
\nonumber\\
&&\quad
\frac{1}{
\Big[q^2\big[q^2+m^2\big]^2+\big\{M^3+\mu
\big[q^2+m^2\big]\big\}^2\Big]}
\,,\label{corr-beforepartial}
\end{eqnarray}
where we have defined
\begin{eqnarray}
f_{\rho}^{-+}(k,q)
&=&
[q\cdot(k-q)]\big[(k-q)^2+m^2\big]^2\big[q^2+m^2\big]^2
+\nonumber\\
&&
+2\,
\big\{M^3+\mu
\big[(k-q)^2+m^2\big]\big\}\big[(k-q)^2+m^2\big]
\big\{M^3+\mu
\big[q^2+m^2\big]\big\}\big[q^2+m^2\big]\,.
\end{eqnarray}
From the knowledge of the poles of the propagator, solutions $y_0=-\omega$ and $y_{\pm}=(-\omega_r\pm i\theta)$ of the following cubic equation
\begin{eqnarray}
y\big[y+m^2\big]^2+\big\{M^3+\mu
\big[y+m^2\big]\big\}^2
&=&0
\,,\label{def-poles}
\end{eqnarray}
we may decompose each propagator appearing in \eqref{corr-beforepartial} as
\begin{eqnarray}
\frac{
1
}{
\Big[q^2\big[q^2+m^2\big]^2+\big\{M^3+\mu
\big[q^2+m^2\big]\big\}^2\Big]}
&=&
\frac{R}{q^2+\omega}
+
\frac{R_+}{q^2+\omega_r+i\theta}
+
\frac{R_-}{q^2+\omega_r-i\theta}
\,,
\label{def-Rs}
\end{eqnarray}
where $R$ and $R_{\pm}$ can be obtained from $\omega, \omega_r$ and $\theta$. In appropriate GeV units, we find
\begin{equation}\label{numbers}
R\approx2.467\,, \omega\approx 0.849\,, R_\pm\approx -1.234\pm i\,15.121\,, \omega_\pm=0.214\pm i\,0.052\,.
\end{equation}
With the previous numbers filled in, we encounter a real pole and a pair of $cc$ poles, corresponding to the $i$-particles of the seminal works \cite{Baulieu:2009ha} where it has been discussed how a pair of such poles can be combined to give a physical, i.e.~consistent with the K\"{a}ll\'{e}n-Lehmann representation, contribution to the bound state propagator. For the remainder of this work, we shall thence only be concerned by this physical part of the correlator, under the assumption that any unphysical piece, originating from combining poles in pairs that are not $cc$, will eventually cancel out when a full-fledged analysis and methodology to deal with $i$-particles would become feasible. Taking this result in eq.~\eqref{corr-beforepartial}, only the terms combining two Yukawa poles ($\omega$) or complex-conjugate poles ($\omega_r\pm i\theta$ with $\omega_r\mp i\theta$) will contribute to the physical part of the spectral function we are interested in. Thus we may write:
\begin{eqnarray}
\Braket{
\rho_{\rho}^-
\rho_{\rho}^+}_k
&=&
8 \int\frac{\d^4q}{(2\pi)^4}
\,
f_{\rho}^{-+}(k,q)\,
\bigg\{
\frac{R^2}{\big[(k-q)^2+\omega\big]\big[q^2+\omega
\big]}
+\frac{R_+R_-}{\big[(k-q)^2+\omega_r+i\theta\big]\big[q^2+\omega_r-i\theta
\big]}
\nonumber\\&&
+\frac{R_-R_+}{\big[(k-q)^2+\omega_r-i\theta\big]\big[q^2+\omega_r+i\theta
\big]}
\bigg\}+\big\{{\rm unphysical} \big\}
\,,\label{corr-afterpartial}
\end{eqnarray}
\subsection{The spectral representation\label{SpecRep-rho}}
In order to search in the next Section for a physical pole associated with the charged $\rho$-meson, we need to obtain the spectral representation of the correlator \eqref{corr-afterpartial}. Following \cite{Dudal:2010wn}, we can use the Cutkosky cut rules and their generalization for the case of complex poles to derive the spectral function $\rho^{-+}(\tau)$:
\begin{eqnarray}
\rho^{-+}(\tau=E^2)&=&\frac{1}{\pi}\, {\rm Im}~\Braket{
\rho_{\rho}^-
\rho_{\rho}^+}_{k=(E,\vec{0})}
\,,
\end{eqnarray}
associated with the correlation function $\Braket{
\rho_{\rho}^-
\rho_{\rho}^+}_k$ given in eq.~\eqref{corr-afterpartial}.

The general result we shall apply to each part of eq.~\eqref{corr-afterpartial} corresponds to Sections 2.1 and 2.2 of \cite{Dudal:2010wn} and is stated as follows\footnote{We use for the moment Minkowski notation. The results however can be continued to Euclidean space and complex-conjugate poles, as discussed in \cite{Dudal:2010wn}.}.
Let
\begin{eqnarray}
\mathcal{F}(k,m_1,m_2)&=&
\int \frac{\d^dq}{(2\pi)^d}
f(k,q)\frac{1}{(k-q)^2-m_1^2}\frac{1}{q^2-m_2^2}
\,,\label{form}
\end{eqnarray}
where $f(k,q)$ is a Lorentz scalar, being thus a function of (and only of) the available scalars: $(k-q)^2=m_1^2,\,q^2=m_2^2,$ and $2q\cdot(k-q)=E^2-m_1^2-m_2^2$. Then
\begin{eqnarray}
{\rm Im}~\mathcal{F}(k=(E,\vec{0}),m_1,m_2)=
\frac{1}{2}
\int \frac{\d^dq}{(2\pi)^{(d-2)}}
f((E,\vec{0}),q)~
\theta(E-q^0)\delta\big[
(E-q^0)^2-\omega_{q,1}^2
\big]\theta(q^0)\delta\big[
(q^0)^2-\omega_{q,2}^2
\big]
\,,
\end{eqnarray}
with $\omega_{q,i}^2\equiv \vec{q}^2+m_i^2$. After carrying out the momentum integrations, we arrive at:
\begin{eqnarray}
{\rm Im}~\mathcal{F}(k=(E,\vec{0}),m_1,m_2)&=&
\frac{1}{8}\,
\frac{1}{2^{d-3}\,\pi^{(d-3)/2}\Gamma((d-1)/2)}
\,
\frac{|\vec{q}_0|^{d-3}}{E}\,
f\big((E,\vec{0}),(\omega_{0,2},\vec{q}_0)\big)
\,,
\label{ImF-Cut}
\end{eqnarray}
where we have defined:
\begin{eqnarray}
|\vec{q}_0|^2
&\equiv&
\frac{(E^2-m_1^2-m_2^2)^2-4m_1^2m_2^2}{4E^2}\,,\quad
\omega_{0,i}^2~\equiv~
|\vec{q}_0|^2+m_i^2
\,,\label{omega0i}
\end{eqnarray}
with $\omega_{0,2}=E-\omega_{0,1}$.

In eq.~\eqref{corr-afterpartial}, we have three physical contributions of the form \eqref{form}, explicitly given in Appendix \ref{contrib}, and dimension $d=4$.
The full spectral function associated with the physical part of the correlator $\Braket{\rho_{\rho}^-\rho_{\rho}^+}_k$ (given in eq.~\eqref{corr-afterpartial}) collects all these results, yielding:
\begin{eqnarray}
\rho_{\rm -+}(\tau)&=&
\theta(\tau-\tau_1)~\frac{1}{\pi}\, {\rm Im}~\mathcal{F}_a(\tau)
+\theta(\tau-\tau_2)~\frac{2}{\pi}\, {\rm Im}~\mathcal{F}_b(\tau)
\,,
\end{eqnarray}
with $ {\rm Im}~\mathcal{F}_a(\tau)$ and $ {\rm Im}~\mathcal{F}_b(\tau)$ given, respectively, in eqs.~\eqref{ImF-res-a} and \eqref{Res-(b)} of Appendix \ref{contrib}.

Putting everything together, we eventually obtain
\begin{eqnarray}\label{corr4}
\braket{\rho_\alpha^-\rho_\alpha^+}_k    &=& \int_0^{\infty} \d\tau \frac{\rho^{-+}(\tau)}{\tau+k^2}+\big\{{\rm unphysical} \big\}\,,\quad \rho^{-+}(\tau)\geq0\,,\nonumber\\
\rho^{-+}(\tau)&=&\theta(\tau-\tau_1)\frac{R^2}{\pi^2}\sqrt{1/4-\omega/\tau}\left[\left(\tau/2-\omega\right)\left(\omega+m^2\right)^4+2\left(M^3+\mu(\omega+M^2)\right)^2\left(\omega+m^2\right)^2\right]
\nonumber\\&&+\theta(\tau-\tau_2)\frac{R_+R_-}{\pi^2}\frac{\sqrt{(\tau-2\omega_r)^2-4(\omega_r^2+\theta^2)}}{\tau}\left[\left(\tau/2-\omega_r\right)\left((\omega_r+m^2)^2+\theta^2\right)^2
\right.\nonumber\\&&\left.+2\left(\left(M^3+\mu(\omega_r+m^2)\right)^2+\mu^2\theta^2\right)\left((\omega_r+m^2)^2+\theta^2\right)\right]
\end{eqnarray}
\noindent with thresholds $\tau_1=4\omega$, $\tau_2=2\omega_r+2\sqrt{\omega_r^2+\theta^2}$. Eq.~\eqref{corr4} is the spectral representation of the single bubble approximation to the $\rho$-correlator.

\subsection{Adding (effective) QCD interactions: bubble resummation for $N\to\infty$ with a contact interaction}
The formation of a true bound state requires that we now take into account the QCD interaction structure. In our model, as in QCD, quarks do not interact directly; the force being mediated by gluons. Since our aim in this Section is to show that the model accommodates the formation of meson bound states from the unphysical (positivity-violating) fundamental degrees of freedom and provide semi-analytic estimates of the meson spectrum of the model,
we shall not undergo the very complicated task of
taking the full gluon interaction into account.

In this first exploration of the physical spectrum of our model, we consider the observation made in e.g.~\cite{Roberts:2011wy} that a gluon contact point interaction can give qualitatively good results. Specifically, it is by now well accepted that the Landau gauge gluon propagator $\mathcal{D}(p^2)$ becomes dynamically massive-like in the infrared. We make the assumption that the gluon is massive-like in the region $p<1$-$2~\text{GeV}$, where the relevant infrared QCD physics is supposed to happen, and that it can be approximated by a constant in momentum space,
\begin{equation}\label{gluonpar}
\mathcal{D}(p^2)=\frac{1}{\Delta^2}\, ,
\end{equation}
or by $\mathcal{D}(x-y)\propto \delta(x-y)$ in position space. Integrating out at lowest order such gluon leads to a NJL-like (contact) interaction between the quarks, more precisely one finds after some Fierz rearranging \cite{Buballa:2003qv} the interaction
\begin{equation}
\frac{1}{2}G (\overline q\vardiamond q)^2 \,+~{\rm lower~ in}~ 1/N\, ,
\end{equation}
where  $\vardiamond\in\{1,i\gamma^5,\frac{i}{\sqrt{2}}\gamma_\mu,\frac{i}{\sqrt{2}}\gamma_\mu\gamma_5\}$ and $G=\frac{2g^2}{\Delta^2}\frac{N^2-1}{N}$. To allow for additional simplification without throwing away the crucial dynamics, we furthermore consider  the leading order in $1/N$.

For the coupling $G$, we estimate an appropriate value as follows. In the Landau gauge, the gluon and ghost propagator form factors, $\mathcal{Z}_{gl}(p^2,\omu^2)$ and $\mathcal{Z}_{gh}(p^2,\omu^2)$, can be combined into a renormalization-scale-independent strong coupling constant,
\begin{equation}
4\pi g^2(p^2)\equiv\alpha(p^2)=\alpha(\omu^2) \mathcal{Z}_{gl}(p^2,\omu^2) \mathcal{Z}_{gh}^2(p^2,\omu^2),
\end{equation}
where  $\omu$ is the renormalization scale, see e.g.~\cite{Fischer:2002hna}. Using the most recent lattice data on this matter \cite{Bornyakov:2013pha}, albeit for $N=2$ without fermions\footnote{See the recent works \cite{Blossier:2013ioa} for a thorough study including multiple dynamical flavours.}, which rely on a MOM scheme ($\mathcal{Z}_{gh}=\mathcal{Z}_{gl}=1$ at $p^2=\omu^2$,  with $\omu=2.2~\text{GeV}$), we can estimate from their Figure.~5 that $\alpha(\omu)\sim0.5$. For the ``constant'' gluon propagator in the MOM scheme, we may simply set\footnote{We remind here that in the MOM scheme at renormalization scale $\omu$, by definition, $\mathcal{D}(p^2=\omu^2)=\frac{1}{\omu^2}$. Given the constant propagator modeling of eq.~\eqref{gluonpar}, it looks natural to set $\Delta^2=\omu^2$ as a kind of mean value for the gluon propagator.} $\Delta^2=\omu^2$, thereby overestimating the UV and underestimating the IR, to roughly approximate the coupling $G$ as:
\begin{equation}
G\sim 8.5~\text{GeV}^{-2}\, .
\end{equation}
Coincidentally, this value falls nicely in the NJL ballpark \cite{Klevansky:1992qe}. We remark here, however, that the NJL parameters are fixed by matching to experimental data for the hadron spectrum, while in the current analysis we do not use \emph{any} such input;  our predicted estimates rely solely on lattice data for confined degrees of freedom, i.e quarks and gluons (or analytical descriptions thereof).  In what follows, we use $G=5,7.5,10~\text{GeV}^{-2}$ as exemplary values.

\begin{figure}[h]
   \centering
%   \vspace{-2cm}
       \includegraphics[width=8cm,angle=0]{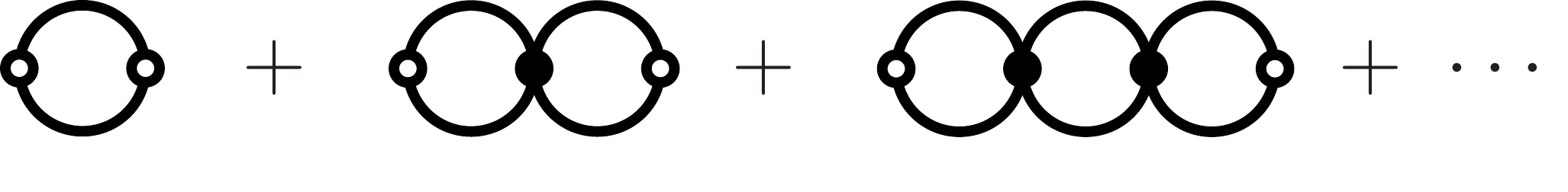}
 %      \vspace{-2cm}
               \caption{Bubble diagrams for the meson correlator. Full (open) circles represent four-fermion vertices (meson operator insertions); solid lines are non-perturbative quark propagators.}\label{bubblechain}
\end{figure}

In the large-$N$ approximation, we can then consistently consider the sum of bubble diagrams\footnote{This is also known as the Random Phase Approximation (see e.g.~\cite{Buballa:2003qv}).} in our quark model, see Figure~\ref{bubblechain}.
The four-fermion coupling includes {\it a priori} all interaction channels $\vardiamond\in\{1,i\gamma^5,\frac{i}{\sqrt{2}}\gamma_\mu,\frac{i}{\sqrt{2}}\gamma_\mu\gamma_5\}$, while the initial and final insertions carry the vectorial character of the $\rho$ meson. In this case, only the vector four-fermion coupling $\gamma_\mu$ contributes. Indeed, bubbles that connect a vector insertion $\gamma^{\mu}$ with a coupling in a different channel (i.e.~$1,\gamma^5,\gamma_\mu\gamma_5$) are absent: they either vanish identically due to Dirac algebra and integration symmetries or produce a term $\propto k^{\mu} f(k^2)$, which does not contribute to the transverse correlator  (cf.~\eqref{corr1}) under consideration (see Appendix \ref{channels} for details).

The bubble chain resummation itself reduces thus to a geometric series involving the one-loop result, \eqref{corr4}. With the already derived spectral form at one loop, eq. \eqref{corr4}, we then get for the   resummed $\rho$-meson   correlator:
\begin{equation}
    \mathcal{R}^{-+}(k^2)= \frac{\mathcal{F}^{-+}(k^2)}{1+\frac{G}{2} \mathcal{F}^{-+}(k^2)}\,,
\end{equation}
where $\mathcal{F}^{-+}(k^2)$ is the physical part of the correlator $\braket{\rho_\alpha^-\rho_\alpha^+}_k $ at one loop:
\begin{equation}
\mathcal{F}^{-+}(k^2)=\int_0^\infty \frac{\rho^{-+}(\tau)\d\tau }{\tau+k^2}\, .
\end{equation}
An important observation is that the branch cut structure of the resumed expression $\mathcal{R}^{-+}(k^2)$ is determined by that of its one-loop counterpart $\mathcal{F}^{-+}(k^2)$, in particular $\mathcal{R}^{-+}(k^2)$ has a physical KL form if $\mathcal{F}^{-+}(k^2)$ does.

\subsection{Existence of the physical bound state pole and results for the mass and decay constant}

The remaining task is to identify whether $\mathcal{R}^{-+}(k^2)$ allows for poles in the physical region, $\max(-\tau_1,-\tau_2)<k^2<0$.  Therefore, we need to solve the gap equation:
\begin{equation}
\mathcal{F}^{-+}(k^2)=-2/G\, .
\end{equation}
As is common for $4d$ quantum field theories, the quantity $\mathcal{F}^{-+}(k^2)$ is divergent due to its violent UV behaviour, explicitly visible from the integral representation \eqref{corr4}. We assume dimensional regularization and Landau gauge renormalization factors to kill off sub-loop divergences, whereas the residual infinities in the composite operator Green function can be taken care of by additive subtractions in the BPHZ approach. Taking $n>0$ subtractions at scale $\mathcal{T}>\max(-\tau_1,-\tau_2)$ corresponds to
\begin{equation}\label{sub1}
    \mathcal{F}^{-+}_{sub}(k^2,\mathcal{T})=(\mathcal{T}-k^2)^n\int_0^{\infty} \d\tau \frac{\rho^{-+}(\tau)}{(\tau+\mathcal{T})^n(\tau+k^2)}\,.
\end{equation}
In the current case $n\geq2$ is required. If no subtractions were to be necessary, then $\mathcal{F}^{-+}(k^2)$ is a strictly decreasing function thanks to $\rho^{-+}(\tau)\geq 0$, so at most one solution is possible. This property is not necessarily maintained at the subtracted level, except for $n=1$. Consequently, spurious extra solutions can appear, caused by the enforced functional behaviour after subtraction. Therefore, we take $n=3$ in which case only a single solution is found.

Next to the pole of the propagator, also the associated residue carries physical information. With the conventions of \cite{Jansen:2009hr,Maris:1999nt}, we define the decay constant of the $\rho$ meson$, f_{\rho^\pm}$, via
\begin{equation}
f_{\rho^\pm} m_{\rho^\pm} \varepsilon_\mu = \left\langle0| \overline u \gamma_\mu d|\rho^+\right\rangle\, ,
\end{equation}
with $\varepsilon_\mu$ the polarization tensor of the $\rho^+$ meson, normalized as $\varepsilon_\mu\cdot\varepsilon_\mu=3$. From the matrix element representation of the KL spectral density, it becomes clear that $3f_{\rho^\pm}^2m_\rho^2$ corresponds to the residue of  $\mathcal{R}^{-+}(k^2)$ at its pole $-m_{\rho^\pm}^2$.

\begin{figure}[h!]
   \centering
       \includegraphics[width=15cm]{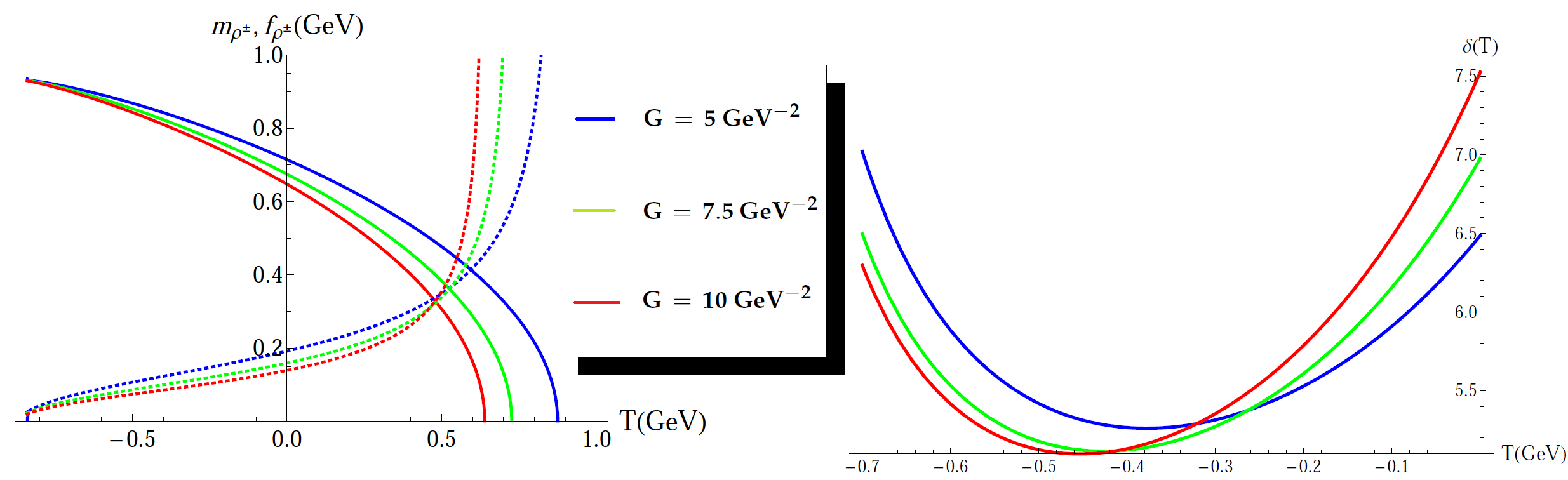}
               \caption{$m_{\rho^\pm}$ (full line) and $f_{\rho^\pm}$  (dashed line) in terms of $\mathcal{T}$ (left) and $\delta(\mathcal{T})$ (right) (see main text for definitions).}
\end{figure}

In Figure~3, we have displayed the estimates obtained for both the $\rho-$meson mass $m_\rho^\pm(\mathcal{T})$ and decay constant $f_{\rho^\pm}(\mathcal{T})$. To get a reasonable value for the subtraction scale $\mathcal{T}$, we rely on the Principle of Minimal Sensitivity (PMS) as observable quantities should not depend on a chosen subtraction scale. Unfortunately, we were not able to find a suitable scale using only $m_{\rho^\pm}(\mathcal{T})$ and its first derivative or $f_{\rho^\pm}(\mathcal{T})$ and its first derivative. The next-to-best PMS strategy that did deliver a reasonable $\mathcal{T}$ was provided for by minimizing a suitable combination of the first derivatives,
 \begin{equation}
  \delta(\mathcal{T})=|\overline m_{\rho^\pm}'(\mathcal{T})|+|\overline f_{\rho^\pm}'(\mathcal{T})|\,,
 \end{equation}
where the prime denotes $\p/\p\mathcal{T}$, and we used the rescaled mass and decay constant, $\overline m_{\rho^\pm}(\mathcal{T})=m_{\rho^\pm}(\mathcal{T})/m_{\rho^\pm}(-\tau_2)$, $\overline f_{\rho^\pm}(\mathcal{T})=f_{\rho^\pm}(\mathcal{T})/f_{\rho^\pm}(-\tau_2)$ to work with dimensionless variables of the same order of magnitude, so that they influence the minimization with similar weight. We choose the smallest possible scale, viz.~$\mathcal{T}=-\tau_2$, as reference. Figure~3 shows that $\delta(\mathcal{T})$ develops a minimum. We find $\mathcal{T}^{G=5,~7.5,~10}_\ast\approx-0.38,-0.43,-0.46$ GeV leading to the following $\rho-$meson mass estimates:
 \begin{equation}
m_{\rho^\pm}^{G=5,~7.5,~10}\approx0.84,0.83,0.83~\text{GeV}\,.
\end{equation}
For the decay constant, we have:
\begin{equation}
f_{\rho^\pm}^{G=5,~7.5,~10}\approx0.13,0.10,0.09~\text{GeV}\, .
\end{equation}
 The mass estimate is essentially stable, while the decay constant seems to be more sensitive to the value of the coupling.
 Even though we are only interested in estimates and will not undergo a detailed error analysis, it is at least reassuring that we find a result not too far off the experimental $\rho$ meson mass, $775.49\pm0.34~\text{MeV}$ \cite{Beringer:1900zz}, despite the approximations made (contact point interaction, bubble approximation), next to having used heavier bare quarks (the current quark mass is fixed by available lattice data for the quark propagator: $\mu=14~$ MeV.). The parameters involved were the 2 vacuum condensates introduced in Section 2 and a value for the coupling $G$, all of them fitted or estimated from lattice data in the quark/gluon/ghost sector. We stress that no empirical input was
used, nor any hadron data, in contrast to what is usually done in analogous estimates in NJL models. For the decay constant, we can quote experimental and lattice estimates, $f_{\rho^\pm}^{exp}\approx 0.208~\text{GeV}$ \cite{Becirevic:2003pn}, $f_{\rho^\pm}^{latt}\approx 0.25~\text{GeV}$ \cite{Jansen:2009hr}. Our results are somewhat lower, though it must be remarked that the heavier the particle mass, the smaller the decay will get for a fixed residue. In fact, if one, for example, sets $\mathcal{T}$ to its value at $G=5~\text{GeV}^2$ where $m_{\rho^\pm}$ equals its experimental value, we obtain $f_{\rho^\pm}\approx 0.16~\text{GeV}$, whose deviation from the experimental value is of the same size as for the lattice result.

The attentive reader might wonder why we did not present a similar resummation analysis for the pion, this to explicitly verify its masslessness in the chiral limit.

We can compare the situation to the NJL model, see e.g.~the review \cite{Klevansky:1992qe}​. There, it is explicitly shown that (at one loop) the pion pole mass remains zero in the chiral limit, in accordance with the earlier current algebra argument.
 
The main reason we did not include such explicit analysis is the following. In the approximation we are working with, we would be tempted to use the nonlocal pion field eq.~\eqref{nonlocalpi} and proceed in analogy with \cite{Klevansky:1992qe}. In this case we would be working with the nonlocal version of the model, after integrating out the auxiliary fields, and we could also stick to using only a quartic $(\overline\psi \psi)^2$ interaction, as for the vector meson. This setup is however not sufficient to guarantee a consistent leading-order estimate of the position of the pole for the pion in our model, which can only be attained through a much more technical procedure. The physical reason for this is the fact that the pion operator in our model is a nonlocal quantity which is affected by interactions. Therefore, any calculation that includes interactions explicitly to a certain order should also adopt a pion field operator that is consistent to the same order. We believe such a technical discussion is out of the scope of the current paper and our general understanding is that, given the current algebra proof that the (nonlocal) pion is massless in our model, this property will anyhow be satisfied if interactions are added in a consistent way (vertices respecting the symmetries; consistent summation of diagrams, etc), in complete analogy with what happens in the NJL example.

\section{Summary and outlook}
We presented a novel quark model that displays quark positivity violation and dynamical chiral symmetry breaking, so that the mesonic spectrum can be described
as composed of bound states of unphysical (positivity-violating) degrees of freedom. This positivity violation can be interpreted as a messenger of confinement, since no observable quark singlets can enter the spectrum. We have described the dynamical origin of the model and showed that its predicted tree-level quark propagator
has an analytical expression (with one real and two {\it cc} poles) that fits well the available lattice data and violates both reflection positivity conditions for the fermionic K\"{a}ll\'{e}n-Lehmann spectral representation.
The formation of mesonic bound states was also addressed and estimates of mass and decay constant of the vector ($\rho$) meson were presented, as an example of the predicting capability of the proposed framework.

The renormalizability of the confining quark model guarantees that its ultraviolet behavior
is equivalent to that of perturbative QCD. On the other hand, the infrared non-perturbative physics is also encoded, since the underlying quark excitations are shown to have a tree level behavior which is compatible with lattice data for the propagator in the Landau gauge, allowing also for mesonic mass estimates. In this sense, the quark degrees of freedom in the model seem to describe well a large momentum range, successfully interpolating between perturbative QCD and the infrared non-perturbative regime of strong interactions.

Moreover, the model has a dynamical origin, inspired by the Gribov-Zwanziger framework for the gluon sector, and has soft BRST breaking as a crucial ingredient to realize an extended consistent scenario of color confinement, in which both gluons and quarks violate positivity conditions and cannot be associated with asymptotic particles. Interestingly, the available soft BRST breaking terms in the quark sector also generate D$\chi$SB, suggesting a possible route to investigate the possibility of confinement and chiral symmetry breaking as related phenomena.

There are many issues that can be analyzed in the near future using the setup presented here. An interesting application that is underway \cite{prep} is the extension of the model to study quark thermodynamics at finite temperature and density. It would of course be interesting to consider as well a full-blown study of the effective action $\Gamma$; a challenging task \cite{Verschelde:1995jj,Knecht:2001cc} which can, however, avoid using lattice input.

\section*{Acknowledgments}
D.~D.~acknowledges support from the Research-Foundation Flanders, S.~P.~S., L.~F.~P.~and M.~S.~G.~from CNPq-Brazil, Faperj, SR2-UERJ and CAPES. L.~F.~P. acknowledges Alexander von Humboldt Foundation and the brazilian program ``Atra\c c\~ ao de Jovens Talentos  -- Ci\^ encia sem Fronteiras'' (grant n. 301111/2014-6) for fellowships at different stages of the current work. We thank O.~Oliveira for Figure~1 and the fit, and U.~Heller for the data of \cite{Parappilly:2005ei}.

\appendix

\section{Positivity conditions from the spectral representation for Dirac fermions\label{ApPos}}
In this   Appendix   we  outline   the derivation of a spectral representation for Dirac fermions, following \cite{ColemanNotes}.
In what follows, we work first in Minkowski space and   afterwards we can extract the Euclidean version  .

The decomposition of the fermion propagator in a complete set of momentum eigenstates yields:
\begin{eqnarray}
\langle 0|
\psi(x)\overline{\psi}(y)
|0\rangle
&=&
\int_0^{\infty} \d a \rho_+(a)(i\slashed{\partial}+a)\Delta(x-y;a^2)
+\int_0^{\infty} \d a \rho_-(a)(i\slashed{\partial}-a)\Delta(x-y;a^2)
\,,\nonumber\\
\end{eqnarray}
where:
\begin{eqnarray}
\Delta(x-y;m^2)
&=&
\int\frac{\d^3p}{(2\pi)^3}\frac{{\rm e}^{-ip\cdot(x-y)}}{2\sqrt{\vec{p}^2+m^2}}\,,
\\
\rho_+(a)\,,
\geq0
\\
\rho_-(a)
\geq 0
\,.
\end{eqnarray}
The positivity of the functions $\rho_{\pm}(a)$ is directly related to the positivity of the norm of fermionic states with  given spin and parity ($J^P=\frac{1}{2}^{\pm}$) in Hilbert space (cf. \cite{ColemanNotes}, for explicit expressions). In perturbation theory, one has $\rho_+(a)=\delta(a-m)$ and $\rho_-(a)=0$.

The Feynman propagator may be constructed from this result, including the time-ordering operator:
\begin{eqnarray}
\langle 0|T
\psi(x)\overline{\psi}(y)
|0\rangle
&=&
\int_0^{\infty} \d a \big[\rho_+(a)(i\slashed{\partial}+a)+
\rho_-(a)(i\slashed{\partial}-a)
\big]
\nonumber\\&&\quad\quad
\Big[
\theta(x^0-y^0)\Delta(x-y;a^2)
+\theta(y^0-x^0)\Delta(y-x;a^2)
\Big]
\,,\nonumber\\
&=&
\int\frac{\d^4p}{(2\pi)^4}
{\rm e}^{-ip\cdot (x-y)}
\int_0^{\infty}\d a \Big(
\rho_+(a)\frac{i}{\slashed{p}-a+i\epsilon}
+
\rho_-(a)
\frac{i}{\slashed{p}+a+i\epsilon}
\Big)
\,,
\end{eqnarray}
where the contributions of  solutions of the Dirac equation with positive and negative energies (i.e. particle and anti-particle) are made explicit and connected to the (positive) spectral functions $\rho_{\pm}$. This result may be rearranged in the standard KL representation in momentum space:
\begin{eqnarray}
\langle 0|T
\psi\overline{\psi}
|0\rangle_p
&=&
\int_0^{\infty}\d a
\frac{i}{p^2-a^2}
\big[
\rho_+(a)(\slashed{p}+a)+
\rho_-(a)(\slashed{p}-a)
\big]
\nonumber\\
&=&
\int_0^{\infty}\d a
\frac{i}{p^2-a^2}
\big(
\slashed{p} [\rho_+(a)+
\rho_-(a)]
+a[\rho_+(a)-
\rho_-(a)]
\big)
\\
&=&
\int_0^{\infty}\d s
\frac{i}{p^2-s}
\left(
\slashed{p} \frac{\rho_+(\sqrt{s})+
\rho_-(\sqrt{s})}{2\sqrt{s}}
+\frac{\rho_+(\sqrt{s})-
\rho_-(\sqrt{s})}{2}
\right)\\
&=&
\int_0^{\infty}\d s
\frac{i}{p^2-s}
\big(
\slashed{p} \rho_v(\sqrt{s})+\rho_s(\sqrt{s})
\big)\label{KL-Mink}
\end{eqnarray}
with the spectral functions satisfying the following relations:
\begin{eqnarray}
\rho_v(a)&=&\frac{\rho_+(a)+\rho_-(a)}{2a}\,,
\\
\rho_s(a)&=&\frac{\rho_+(a)-\rho_-(a)}{2}\,.
\end{eqnarray}
Using $\rho_{\pm}\geq0$, the positivity conditions become:
\begin{eqnarray}
\rho_v(a)\geq0\,,
\nonumber\\
a\rho_v(a)-\rho_s(a) =\rho_-(a)\geq0\,.
\label{PosCond}
\end{eqnarray}
Finally, for the propagator in momentum space we may write:
\begin{eqnarray}
\langle 0|T
\psi\overline{\psi}
|0\rangle_p
&=&
\slashed{p} \sigma_v(p^2)+\sigma_s(p^2)\,,\\
\sigma_v(p^2)&=&
\int_0^{\infty}\d s
\frac{i}{p^2-s}
\rho_v(\sqrt{s})\,,
\\
\sigma_s(p^2)&=&
\int_0^{\infty}\d s
\frac{i}{p^2-s}
\rho_s(\sqrt{s})\,,
\\
\sigma_-(p^2)&=&
\int_0^{\infty}\d s
\frac{i}{p^2-s}
\big[\sqrt{s}\rho_v(\sqrt{s})-\rho_s(\sqrt{s})
\big]
\end{eqnarray}
where $\sigma_v$ and $\sigma_-$ are the scalar functions extracted from the propagator that should satisfy the Osterwalder-Schrader axiom. Note that $\sigma_s$  does not need to be positive .
It is also interesting to realize that the simple assumption of existence of a asymptotic Dirac particle/anti-particle leads directly to a KL representation for the corresponding two-point function. So, in this standard analysis, the existence of a KL representation is directly built in.

From the form of the free propagators in Minkowski and Euclidean space,
\begin{eqnarray}
\Delta_M(\slashed{p})&=&\frac{i}{\slashed{p}-m}\,,
\\
\Delta_E(\slashed{p})&=&\frac{1}{i\slashed{p}+m}=\frac{-i\slashed{p}+m}{p^2+m^2}\,,
\end{eqnarray}
one infers directly the relation between these quantities: $\Delta_E(\slashed{p})=i\Delta_M(-i\slashed{p})$. In the KL representation eq. \eqref{KL-Mink}, this same transformation implies that the positive spectral functions remain the same in Euclidean space:
\begin{eqnarray}
\Delta_E(\slashed{p})
&=&
\int_0^{\infty}\d s
\frac{1}{p^2+s}
\big(
-i\slashed{p} \rho_v(\sqrt{s})+\rho_s(\sqrt{s})
\big)
=
-i\slashed{p} \sigma^E_v(p^2)+\sigma^E_s(p^2)
\,,
\\\nonumber\\
&&\rho_v\geq 0\,,\\
&&
a\rho_v(a)-\rho_s(a)=\rho_-(a)\geq0 \quad \Leftrightarrow \quad  a\rho_v(a)\geq\rho_s(a)\,,\label{Pos2}
\end{eqnarray}
with the definitions:
\begin{eqnarray}
\sigma_v(p^2)&=&
\int_0^{\infty}\d s
\frac{1}{p^2+s}
\rho_v(\sqrt{s})\,,
\\
\sigma_s(p^2)&=&
\int_0^{\infty}\d s
\frac{1}{p^2+s}
\rho_s(\sqrt{s})\,.
\end{eqnarray}
Inspection also reveals that the KL representation always follows the form of the free propagator of the theory, which is nice for bookkeeping.

Furthermore, going back to the Osterwalder-Schrader axiom of reflection positivity, one may identify the scalar functions extracted from the fermion propagator that should enter the (positive) Schwinger function in eq.\eqref{posSF} as $\sigma_v(p^2)$ and
\begin{eqnarray}
\sigma_-(p^2)
&=&
\int_0^{\infty}\d s
\frac{1}{p^2+s}
\big[\sqrt{s}\rho_v(\sqrt{s})-\rho_s(\sqrt{s})
\big]\,.
\end{eqnarray}
It is however more convenient to look at the second positivity condition, eq.\eqref{Pos2}, through the last inequality, namely: $a\rho_v(a)\geq \rho_s(a)$.
This relation gives a straightforward relation between the corresponding Schwinger Functions, since in Euclidean space they are related by a Laplace Transform, which in turn does not affect the inequality. Indeed:
\begin{eqnarray}
&&a\rho_v(a)\geq \rho_s(a)
\\
&&\Rightarrow \int_0^{\infty}\d a a\rho_v(a){\rm e}^{-ta}\geq \int_0^{\infty}\d a\rho_s(a){\rm e}^{-ta}
\\
&&\Rightarrow -\partial_t \int_0^{\infty}\d a \rho_v(a){\rm e}^{-ta}\geq \int_0^{\infty}\d a\rho_s(a){\rm e}^{-ta}
\\\nonumber\\
\Rightarrow -\partial_t \Delta_v(t)\geq \Delta_{s}(t)
\end{eqnarray}
where $\Delta_{I}(t)$ are the Schwinger functions:
\begin{eqnarray}
\Delta_{I}(t)&=&\frac{1}{2\pi}\int_{-\infty}^{\infty}\d p {\rm e}^{ipt}\sigma_I(p^2)\,.
\end{eqnarray}
In the last line we have used the fact that in Euclidean space the Schwinger function obeys:
\begin{eqnarray}
\Delta_{I}(t)&=&\frac{1}{2\pi}\int_{-\infty}^{\infty}\d p {\rm e}^{ipt}\sigma_I(p^2)
\\
&=&\frac{1}{2\pi}\int_{-\infty}^{\infty}\d p {\rm e}^{ipt}\int_0^{\infty}\d s \frac{\rho_I(\sqrt{s})}{p^2+s}
\\
&=&\int_0^{\infty}\d a \rho_I(a) {\rm e}^{-at}\,.
\end{eqnarray}
Note that, in writing the Schwinger function as a Laplace transform of the spectral function, the existence of a KL representation is assumed.

\section{Imaginary parts contributing to the spectral representation of the $\rho$ meson\label{contrib}}
In this Appendix we write down explicitly the three physical contributions, of the form \eqref{form}, to eq.~\eqref{corr-afterpartial} in dimension $d=4$.
%In eq.~\eqref{corr-afterpartial}, we have three physical contributions of the form \eqref{form}, explicitly given in Appendix \ref{contrib}, and dimension $d=4$.
\paragraph{\underline{$m_1^2=m_2^2=\omega$ and $f=8f_{\rho}^{-+}R^2$:}}
\begin{eqnarray}
{\rm Im}~\mathcal{F}_a(E^2)&=&
\frac{1}{8\pi}\,
\frac{|\vec{q}_a|}{E}\,
8R^2\,f_{\rho}^{-+}\big((E,\vec{0}),(\omega_{a,2},\vec{q}_a)\big)
\,,\end{eqnarray}
with
\begin{eqnarray}
|\vec{q}_a|^2
&\equiv&
\frac{E^2}{4}-\omega\,,\quad
\omega_{a,i}^2~\equiv~
\frac{E^2}{4}
\,,
\end{eqnarray}
and $(k-q)^2=m_1^2\mapsto\omega,\,q^2=m_2^2\mapsto\omega,$ and $2q\cdot(k-q)=E^2-m_1^2-m_2^2\mapsto E^2-2\omega$, so that
\begin{eqnarray}
f_{\rho}^{-+}\big((E,\vec{0}),(\omega_{a,2},\vec{q}_a)\big)
&=&
[E^2/2-\omega]\big[\omega+m^2\big]^4+2\,
\big\{M^3+\mu
\big[\omega+m^2\big]\big\}^2\big[\omega+m^2\big]^2\,.
\end{eqnarray}
The final form of this contribution is:
\begin{eqnarray}
{\rm Im}~\mathcal{F}_a(E^2)&=&
\frac{1}{\pi}\,
\frac{\sqrt{E^2/4-\omega}}{E}\,
R^2\,
\Big\{
[E^2/2-\omega]\big[\omega+m^2\big]^4
+2\,
\big\{M^3+\mu
\big[\omega+m^2\big]\big\}^2\big[\omega+m^2\big]^2
\Big\}
\,,
\label{ImF-res-a}\nonumber\\
\end{eqnarray}
and the threshold for multiparticle production (related to $\tau_0$, the lower limit of integration of the spectral representation) is $\tau_1=4\omega$.

\paragraph{\underline{$m_1^2=\omega_r+i\theta,\,m_2^2=\omega_r-i\theta$ and $f=8f_{\rho}^{-+}R_+R_-$:}}
\begin{eqnarray}
{\rm Im}~\mathcal{F}_b(E^2)&=&
\frac{1}{8\pi}\,
\frac{|\vec{q}_b|}{E}\,8
R_+R_-\,f_{\rho}^{-+}\big((E,\vec{0}),(\omega_{b,2}=\omega_{b,-},\vec{q}_b)\big)
\,,\nonumber\\
\end{eqnarray}
with
\begin{eqnarray}
|\vec{q}_b|^2
&\equiv&
\frac{(E^2-2\omega_r)^2-4(\omega_r^2+\theta^2)}{4E^2}
\,,\quad \omega_{b,\pm}^2~\equiv~
\frac{(E^2-2\omega_r)^2-4(\omega_r^2+\theta^2)}{4E^2}+\omega_r\pm i\theta
\,,
\end{eqnarray}
and $(k-q)^2=m_1^2\mapsto\omega_r+i\theta,\,q^2=m_2^2\mapsto\omega_r-i\theta,$ and $2q\cdot(k-q)=E^2-m_1^2-m_2^2\mapsto E^2-2\omega_r$, so that
\begin{eqnarray}
f_{\rho}^{-+}\big((E,\vec{0}),(\omega_{b,-},\vec{q}_b)\big)
&=&
[E^2/2-\omega_r]\big[(\omega_r+m^2)^2+\theta^2\big]^2
+\nonumber\\
&&
+2\,
\big\{\big[M^3+\mu
(\omega_r+m^2)\big]^2+\mu^2\theta^2\big\}\, \big[(\omega_r+m^2)^2+\theta^2\big]\,.
\end{eqnarray}
The final form of this contribution is:
\begin{eqnarray}
{\rm Im}~\mathcal{F}_b(E^2)&=&
\frac{1}{\pi}\,
\frac{\sqrt{(E^2-2\omega_r)^2-4(\omega_r^2+\theta^2)}}{2E^2}
\,
R_+R_-\,\times
\nonumber\\&&
\!\!\!\!\!\!\!\!\times~
\Big\{[E^2/2-\omega_r]\big[(\omega_r+m^2)^2+\theta^2\big]^2
+2\,
\big\{\big[M^3+\mu
(\omega_r+m^2)\big]^2+\mu^2\theta^2\big\}\, \big[(\omega_r+m^2)^2+\theta^2\big]
\Big\}\label{Res-(b)}
\end{eqnarray}
and the threshold is in this case $\tau_2=2\omega_r+2\sqrt{\omega_r^2+\theta^2}$, below which the square root above becomes imaginary, signaling the instability related to multiparticle production.

\paragraph{\underline{$m_1^2=\omega_r-i\theta,\,m_2^2=\omega_r+i\theta$ and $f=f_{\rho}^{-+}R_+R_-$:}} it is straightforward to see that this contribution will be exactly the same as in $(b)$, eq.~\eqref{Res-(b)}.

\section{Absence of channel mixing in the bubble chain resummation \label{channels}}

 Let us show that bubbles that connect a vector insertion $\gamma^{\mu}$ with a coupling in a different channel (i.e.~$1,\gamma^5,\gamma_\mu\gamma_5$) are absent: they either vanish identically due to Dirac algebra and integration symmetries or produce a term $\propto k^{\mu} f(k^2)$, which does not contribute to the transverse correlator  (cf.~\eqref{corr1}) under consideration. Indeed, the basic integral
\begin{eqnarray}
\mathcal{I}_{\mu\vardiamond}(k)
=\int\frac{\d^4p}{(2\pi)^4}
\frac{-{\rm Tr}\left[(i\slashed{p}+m)\gamma_{\mu}(i\slashed{p}-i\slashed{k}+m)\vardiamond\right]
}{[p^2+m^2][(p-k)^2+m^2]}
\label{Idiamond}
\end{eqnarray}
furnishes the following trivial results for nonvector $\vardiamond$-insertions, $\vardiamond\in  \{ 1,\gamma^5,\frac{1}{\sqrt{2}}\gamma_{\mu}\gamma^5 \} $:
\begin{itemize}
\item $\vardiamond=1$: in this case, the Dirac trace in the integrand of eq.~(\ref{Idiamond}) reduces to
\begin{eqnarray}
\mathcal{I}_{\mu -}(k)&=&
\int\frac{\d^4p}{(2\pi)^4}
\frac{-{\rm Tr}\left[(i\slashed{p}+m)\gamma_{\mu}(i\slashed{p}-i\slashed{k}+m)\right]
}{[p^2+m^2][(p-k)^2+m^2]}\nonumber
\\
&=&
-4im
\int\frac{\d^4p}{(2\pi)^4}
\frac{2p_{\mu}-k_{\mu}
}{[p^2+m^2][(p-k)^2+m^2]}~\propto~k_{\mu} \, f(k^2)
\,,
\end{eqnarray}
where the last line is obtained after the introduction of a Feynman parameter in the usual way. Recalling that we are computing a two-point correlation function which is transverse, cfr.~\eqref{corr1}, it is straightforward to conclude that this term $\propto k_{\mu}$ does not contribute to our observable.

\item $\vardiamond=\gamma^5$: the Dirac trace in the integrand of eq.~\eqref{Idiamond} vanishes identically.
\begin{eqnarray}
\mathcal{I}_{\mu 5}(k)&=&
\int\frac{\d^4p}{(2\pi)^4}
\frac{-{\rm Tr}\left[(i\slashed{p}+m)\gamma_{\mu}(i\slashed{p}-i\slashed{k}+m)\gamma^5\right]
}{[p^2+m^2][(p-k)^2+m^2]}~=~0
\,,
\end{eqnarray}

\item $\vardiamond=\frac{1}{\sqrt{2}}\gamma_{\rho}\gamma^5$: in this case, only the term with four gamma matrices survives:
\begin{eqnarray}
\mathcal{I}_{\mu \rho5}(k)&=&\frac{1}{\sqrt{2}}
\int\frac{\d^4p}{(2\pi)^4}
\frac{-{\rm Tr}\left[(i\slashed{p}+m)\gamma_{\mu}(i\slashed{p}-i\slashed{k}+m)\gamma_{\rho}\gamma^5\right]
}{[p^2+m^2][(p-k)^2+m^2]}\nonumber
\\
&=&
4i \frac{1}{\sqrt{2}}
\int\frac{\d^4p}{(2\pi)^4}
\frac{p_{\sigma}(p_{\nu}-k_{\nu})\epsilon_{\sigma\mu\nu\rho}
}{[p^2+m^2][(p-k)^2+m^2]}~=~
-
4i \frac{1}{\sqrt{2}}
\, k_{\nu}\epsilon_{\sigma\mu\nu\rho}\,
k_{\sigma} f(k^2)~=~0\,,
\end{eqnarray}
where we have used the antisymmetry of the $\epsilon$-tensor, while we introduced again a Feynman parameter in the last step.
\end{itemize}


\begin{thebibliography}{99}
\bibitem{McLerran:2007qj}
L.~McLerran and R.~D.~Pisarski, Nucl.\ Phys.\ A {\bf 796} (2007) 83.

\bibitem{Kojo:2009ha}
T.~Kojo, Y.~Hidaka, L.~McLerran and R.~D.~Pisarski, Nucl.\ Phys.\ A {\bf 843} (2010) 37.

\bibitem{Creutz:2011hy}
M.~Creutz, Acta Phys.\ Slov.\  {\bf 61} (2011) 1.

\bibitem{Aoki:2006br}
Y.~Aoki, Z.~Fodor, S.~D.~Katz and K.~K.~Szabo, Phys.\ Lett.\ B {\bf 643} (2006) 46; Y.~Aoki, S.~Borsanyi, S.~Durr, Z.~Fodor, S.~D.~Katz, S.~Krieg and K.~K.~Szabo, JHEP {\bf 0906} (2009) 088.

\bibitem{Bashir:2013zha}
A.~Bashir, A.~Raya, J.~Rodriguez-Quintero,   Phys.\ Rev.\ D {\bf 88} (2013) 054003.

\bibitem{Fontoura:2012mz}
C.~E.~Fontoura, G.~Krein and V.~E.~Vizcarra, Phys.\ Rev.\ C {\bf 87}, 2 (2013) 025206.

\bibitem{Li:2012ay}
D.~Li, M.~Huang and Q.~S.~Yan, Eur.\ Phys.\ J.\ C {\bf 73} (2013) 2615.

\bibitem{Cornwall:2010ap}
J.~M.~Cornwall, Phys.\ Rev.\ D {\bf 83} (2011) 076001; A.~Doff, F.~A.~Machado and A.~A.~Natale, Annals Phys.\  {\bf 327} (2012) 1030.

\bibitem{Fukushima:2003fw}
K.~Fukushima, Phys.\ Lett.\ B {\bf 591} (2004) 277.

\bibitem{Mocsy:2004yt}
A.~Mocsy, F.~Sannino and K.~Tuominen, hep-ph/0401149;  Phys.\ Rev.\ Lett.\  {\bf 92}  (2004) 182302; E.~Fraga and A.~Mocsy, Braz.\ J.\ Phys.\  {\bf 37} (2007) 281.

\bibitem{Megias:2004hj}
E.~Megias, E.~Ruiz Arriola and L.~L.~Salcedo, Phys.\ Rev.\ D {\bf 74}  (2006) 065005.

\bibitem{Schaefer:2007pw}
B.~J.~Schaefer, J.~M.~Pawlowski and J.~Wambach, Phys.\ Rev.\ D {\bf 76} (2007) 074023.

\bibitem{Contrera:2010kz}
G.~A.~Contrera, M.~Orsaria and N.~N.~Scoccola, Phys.\ Rev.\ D {\bf 82} (2010) 054026; K.~I.~Kondo,  Phys.\ Rev.\ D {\bf 82} (2010) 065024.

\bibitem{Hell:2008cc}
T.~Hell, S.~Roessner, M.~Cristoforetti and W.~Weise, Phys.\ Rev.\ D {\bf 79} (2009) 014022.

\bibitem{Herbst:2010rf}
T.~K.~Herbst, J.~M.~Pawlowski and B.~J.~Schaefer, Phys.\ Lett.\ B {\bf 696} (2011) 58.

\bibitem{Parappilly:2005ei}
M.~B.~Parappilly, P.~O.~Bowman, U.~M.~Heller, D.~B.~Leinweber, A.~G.~Williams, J.~B.~Zhang, Phys.\ Rev.\ D {\bf 73} (2006) 054504.

\bibitem{Furui:2006ks}
S.~Furui, H.~Nakajima,   Phys.\ Rev.\ D {\bf 73} (2006) 074503.

\bibitem{Burgio:2012ph}
G.~Burgio, M.~Schrock, H.~Reinhardt, M.~Quandt, Phys.\ Rev.\ D {\bf 86} (2012) 014506.

\bibitem{Bowler:1994ir}
R.~D.~Bowler and M.~C.~Birse, Nucl.\ Phys.\ A {\bf 582} (1995) 655.

\bibitem{Plant:1997jr}
R.~S.~Plant and M.~C.~Birse, Nucl.\ Phys.\ A {\bf 628} (1998) 607.

\bibitem{Contrera:2007wu}
G.~A.~Contrera, D.~Gomez Dumm and N.~N.~Scoccola, Phys.\ Lett.\ B {\bf 661} (2008) 113.

\bibitem{Loewe:2013zaa}
M.~Loewe, F.~Marquez and C.~Villavicencio, Phys.\ Rev.\ D {\bf 88} (2013) 056004.

\bibitem{Marquez:2014kla}
F.~Marquez, Phys.\ Rev.\ D {\bf 89} (2014) 7,  076010.

\bibitem{Marquez:2015bca}
F.~Marquez, A.~Ahmad, M.~Buballa and A.~Raya, Phys.\ Lett.\ B {\bf 747} (2015) 529.

\bibitem{Buballa:1992sz}
M.~Buballa and S.~Krewald, Phys.\ Lett.\ B {\bf 294} (1992) 19.

\bibitem{Gribov:1977wm}
V.~N.~Gribov, Nucl.\ Phys.\ B {\bf 139} (1978) 1.

\bibitem{Vandersickel:2012tz}
N.~Vandersickel, D.~Zwanziger, Phys.\ Rept.\  {\bf 520} (2012) 175; D.~Dudal, J.~A.~Gracey, S.~P.~Sorella, N.~Vandersickel, H.~Verschelde, Phys.\ Rev.\ D {\bf 78} (2008) 065047;   J.~A.~Gracey, Phys.\ Rev.\ D {\bf 82} (2010) 085032; F.~Canfora, L.~Rosa, Phys.\ Rev.\ D {\bf 88} (2013) 045025.

\bibitem{Dudal:2007cw}
D.~Dudal, S.~P.~Sorella, N.~Vandersickel and H.~Verschelde, Phys.\ Rev.\ D {\bf 77} (2008) 071501;  D.~Dudal, J.~A.~Gracey, S.~P.~Sorella, N.~Vandersickel and H.~Verschelde, Phys.\ Rev.\ D {\bf 78} (2008) 065047; D.~Dudal, S.~P.~Sorella and N.~Vandersickel, Phys.\ Rev.\ D {\bf 84} (2011) 065039.

\bibitem{Dudal:2010tf}
D.~Dudal, O.~Oliveira and N.~Vandersickel, Phys.\ Rev.\ D {\bf 81} (2010) 074505; A.~Cucchieri, D.~Dudal, T.~Mendes and N.~Vandersickel,
Phys.\ Rev.\ D {\bf 85} (2012) 094513; D.~Dudal, O.~Oliveira and J.~Rodriguez-Quintero, Phys.\ Rev.\ D {\bf 86} (2012) 105005.

\bibitem{Dudal:2013yva}
D.~Dudal, O.~Oliveira and P.~J.~Silva, Phys.\ Rev.\ D {\bf 89} (2014) 014010.

\bibitem{Dudal:2010cd}
D.~Dudal, M.~S.~Guimaraes and S.~P.~Sorella, Phys.\ Rev.\ Lett.\  {\bf 106} (2011) 062003; Phys.\ Lett.\ B {\bf 732} (2014) 247.

\bibitem{Cucchieri:2014via}
A.~Cucchieri, D.~Dudal, T.~Mendes and N.~Vandersickel, Phys.\ Rev.\ D {\bf 90} (2014) 051501.

\bibitem{Alkofer:2003jj}
R.~Alkofer, W.~Detmold, C.~S.~Fischer and P.~Maris, Phys.\ Rev.\ D {\bf 70} (2004) 014014.

\bibitem{Fukushima:2012qa}
K.~Fukushima, K.~Kashiwa, Phys.\ Lett.\ B {\bf 723} (2013) 360.

\bibitem{benic}
S.~Benic, D.~Blaschke, M.~Buballa, Phys.\ Rev.\ D {\bf 86} (2012) 074002.

\bibitem{Benic:2013eqa}
S.~Benic, D.~Blaschke, G.~A.~Contrera and D.~Horvatic, Phys.\ Rev.\ D {\bf 89} (2014) 016007.

\bibitem{Fukushima:2013xsa}
K.~Fukushima and N.~Su,  Phys.\ Rev.\ D {\bf 88} (2013) 076008.

\bibitem{Su:2014rma}
N.~Su and K.~Tywoniuk,   Phys.\ Rev.\ Lett.\  {\bf 114} (2015) 16,  161601.

\bibitem{prep}
M.~S.~Guimaraes, B.~W.~Mintz and L.~F.~Palhares, Phys.\ Rev.\ D {\bf 92} (2015) 8,  085029.

\bibitem{Capri:2012hh}
M.~A.~L.~Capri, D.~Dudal, M.~S.~Guimaraes, L.~F.~Palhares, S.~P.~Sorella, Int.\ J.\ Mod.\ Phys.\ A {\bf 28} (2013) 1350034.

\bibitem{Baulieu:2009ha}
L.~Baulieu, D.~Dudal, M.~S.~Guimaraes, M.~Q.~Huber, S.~P.~Sorella, N.~Vandersickel, D.~Zwanziger, Phys.\ Rev.\ D {\bf 82} (2010) 025021; D.~Dudal, M.~S.~Guimaraes, S.~P.~Sorella, Phys.\ Rev.\ Lett.\  {\bf 106} (2011) 062003.

\bibitem{Dudal:2013wja}
D.~Dudal, M.~S.~Guimaraes and S.~P.~Sorella, Phys.\ Lett.\ B {\bf 732} (2014) 247.

\bibitem{Capri:2014bsa}
  M.~A.~L.~Capri, M.~S.~Guimaraes, I.~F.~Justo, L.~F.~Palhares and S.~P.~Sorella,   Phys.\ Rev.\ D {\bf 90} (2014) 8, 085010.

\bibitem{Capri:2014fsa}
M.~A.~L.~Capri, D.~Fiorentini and S.~P.~Sorella, Annals Phys.\  {\bf 356} (2015) 320.

\bibitem{Dudal:2009xh}
N.~Maggiore and M.~Schaden, Phys.\ Rev.\ D {\bf 50} (1994) 6616; D.~Dudal, S.~P.~Sorella, N.~Vandersickel and H.~Verschelde, Phys.\ Rev.\ D {\bf 79} (2009) 121701; S.~P.~Sorella, Phys.\ Rev.\ D {\bf 80} (2009);  D.~Dudal and N.~Vandersickel, Phys.\ Lett.\ B {\bf 700} (2011) 369; M.~A.~L.~Capri, A.~J.~Gomez, M.~S.~Guimaraes, V.~E.~R.~Lemes, S.~P.~Sorella and D.~G.~Tedesco, Phys.\ Rev.\ D {\bf 82} (2010) 105019; M.~A.~L.~Capri, A.~J.~Gomez, M.~S.~Guimaraes, V.~E.~R.~Lemes, S.~P.~Sorella and D.~G.~Tedesco, Phys.\ Rev.\ D {\bf 83} (2011) 105001; D.~Dudal and S.~P.~Sorella, Phys.\ Rev.\ D {\bf 86} (2012) 045005; M.~A.~L.~Capri, D.~Dudal, M.~S.~Guimaraes, I.~F.~Justo, L.~F.~Palhares and S.~P.~Sorella, Annals Phys.\  {\bf 339} (2013) 344; P.~Lavrov, O.~Lechtenfeld and A.~Reshetnyak, JHEP {\bf 1110} (2011) 043; A.~Reshetnyak, arXiv:1312.2092 [hep-th]; P.~Y.~Moshin and A.~A.~Reshetnyak, Nucl.\ Phys.\ C {\bf 888} (2014) 92.

\bibitem{Serreau:2012cg}
J.~Serreau, M.~Tissier, Phys.\ Lett.\ B {\bf 712} (2012) 97;   M.~Pelaez, M.~Tissier and N.~Wschebor, Phys.\ Rev.\ D {\bf 88} (2013) 125003; U.~Reinosa, J.~Serreau, M.~Tissier and N.~Wschebor, Phys.\ Rev.\ D {\bf 89} (2014) 105016.

\bibitem{Verschelde:2001ia}
H.~Verschelde, K.~Knecht, K.~Van Acoleyen and M.~Vanderkelen, Phys.\ Lett.\ B {\bf 516} (2001) 307; D.~Dudal, R.~F.~Sobreiro, S.~P.~Sorella and H.~Verschelde, Phys.\ Rev.\ D {\bf 72} (2005) 014016.

\bibitem{Kaplan:2013dca}
D.~B.~Kaplan, arXiv:1306.5818 [nucl-th].

\bibitem{Capri:2015ixa}
M.~A.~L.~Capri {\it et al.}, Phys.\ Rev.\ D {\bf 92} (2015) 4,  045039

\bibitem{Capri:2015pfa}
M.~A.~L.~Capri, D.~Fiorentini and S.~P.~Sorella, Phys.\ Lett.\ B {\bf 751} (2015) 262.

\bibitem{Verschelde:1995jj}
H.~Verschelde, Phys.\ Lett.\ B {\bf 351} (1995) 242.

\bibitem{Knecht:2001cc}
K.~Knecht and H.~Verschelde, Phys.\ Rev.\ D {\bf 64} (2001) 085006.

%\bibitem{workinprogress}
%M.~A.~L.~Capri et al, \emph{work in progress}.

\bibitem{Banks:1975zw}
T.~Banks, S.~Raby, Phys.\ Rev.\ D {\bf 14} (1976) 2182.

\bibitem{Stevenson:1981vj}
P.~M.~Stevenson, Phys.\ Rev.\ D {\bf 23} (1981) 2916.

\bibitem{Jackiw:1974cv}
R.~Jackiw, Phys.\ Rev.\ D {\bf 9} (1974) 1686.

\bibitem{Rojas:2013tza}
E.~Rojas, J.~P.~B.~C.~de Melo, B.~El-Bennich, O.~Oliveira, T.~Frederico, JHEP {\bf 1310} (2013) 193.

\bibitem{Baulieu:2009xr}
L.~Baulieu, M.~A.~L.~Capri, A.~J.~Gomez, V.~E.~R.~Lemes, R.~F.~Sobreiro, S.~P.~Sorella, Eur.\ Phys.\ J.\ C {\bf 66} (2010) 451; L.~Baulieu, S.~P.~Sorella, Phys.\ Lett.\ B {\bf 671} (2009) 481.

\bibitem{Diakonov:1985eg}
D.~Diakonov, V.~Y.~Petrov, Nucl.\ Phys.\ B {\bf 272} (1986) 457.

\bibitem{Diakonov:1987ty}
D.~Diakonov, V.~Y~.Petrov, P.~V.~Pobylitsa, Nucl.\ Phys.\ B {\bf 306} (1988) 809.

\bibitem{Bhagwat:2002tx}
M.~Bhagwat, M.~A.~Pichowsky and P.~C.~Tandy, Phys.\ Rev.\ D {\bf 67} (2003) 054019.

\bibitem{Bhagwat:2003vw}
M.~S.~Bhagwat, M.~A.~Pichowsky, C.~D.~Roberts, P.~C.~Tandy,  Phys.\ Rev.\ C {\bf 68} (2003) 015203.

\bibitem{Pokorski:1987ed}
S.~Pokorski, \emph{Gauge Field Theories}, Cambridge, UK: Univ.~Pr. (1987).

\bibitem{Capri:2014xea}
M.~A.~L.~Capri, D.~R.~Granado, M.~S.~Guimaraes, I.~F.~Justo, L.~F.~Palhares, S.~P.~Sorella and D.~Vercauteren,  Eur.\ Phys.\ J.\ C {\bf 74} (2014) 2961.

\bibitem{Fujikawa:1979ay}
K.~Fujikawa, Phys.\ Rev.\ Lett.\  {\bf 42} (1979) 1195; Phys.\ Rev.\ D {\bf 21} (1980) 2848.

\bibitem{Greensite:2011zz}
J.~Greensite, Lect.\ Notes Phys.\  {\bf 821} (2011) 1.

\bibitem{Alkofer:2006gz}
R.~Alkofer, C.~S.~Fischer and F.~J.~Llanes-Estrada, Mod.\ Phys.\ Lett.\ A {\bf 23} (2008) 1105.

\bibitem{Canfora:2015yia}
F.~E.~Canfora, D.~Dudal, I.~F.~Justo, P.~Pais, L.~Rosa and D.~Vercauteren, Eur.\ Phys.\ J.\ C {\bf 75} (2015) 7,  326.

\bibitem{Krein:1990sf}
G.~Krein, C.~D.~Roberts and A.~G.~Williams, Int.\ J.\ Mod.\ Phys.\ A {\bf 7} (1992) 5607.

\bibitem{Roberts:2007ji}
C.~D.~Roberts, Prog.\ Part.\ Nucl.\ Phys.\  {\bf 61} (2008) 50.

\bibitem{Alkofer:2008tt}
R.~Alkofer, C.~S.~Fischer, F.~J.~Llanes-Estrada, K.~Schwenzer, Annals Phys.\  {\bf 324} (2009) 106.

\bibitem{Osterwalder:1973dx}
   K.~Osterwalder and R.~Schrader, Commun.\ Math.\ Phys.\  {\bf 31} (1973) 83.

\bibitem{Osterwalder:1974tc}
   K.~Osterwalder and R.~Schrader, Commun.\ Math.\ Phys.\  {\bf 42} (1975) 281.

\bibitem{ColemanNotes}
S.~Coleman, arXiv:1110.5013 [physics.ed-ph].

\bibitem{Dudal:2010wn}
D.~Dudal, M.~S.~Guimaraes, Phys.\ Rev.\ D {\bf 83} (2011) 045013.

\bibitem{Roberts:2011wy}
H.~L.~L.~Roberts, A.~Bashir, L.~X.~Gutierrez-Guerrero, C.~D.~Roberts, D.~J.~Wilson, Phys.\ Rev.\ C {\bf 83} (2011) 065206.

\bibitem{Buballa:2003qv}
M.~Buballa, Phys.\ Rept.\  {\bf 407} (2005) 205.

\bibitem{Fischer:2002hna}
C.~S.~Fischer, R.~Alkofer, Phys.\ Lett.\ B {\bf 536} (2002) 177;    A.~C.~Aguilar, D.~Binosi, J.~Papavassiliou, J.~Rodriguez-Quintero, Phys.\ Rev.\ D {\bf 80} (2009) 085018.

\bibitem{Bornyakov:2013pha}
V.~G.~Bornyakov, E.~-M.~Ilgenfritz, C.~Litwinski, V.~K.~Mitrjushkin, M.~Muller-Preussker, arXiv:1302.5943 [hep-lat].

\bibitem{Blossier:2013ioa}
B.~Blossier, P.~Boucaud, M.~Brinet, F.~De Soto, V.~Morenas, O.~Pene, K.~Petrov, J.~Rodriguez-Quintero,   Phys.\ Rev.\ D {\bf 89} (2014) 1,  014507; P.~Boucaud, M.~Brinet, F.~De Soto, V.~Morenas, O.~Pene, K.~Petrov, J.~Rodriguez-Quintero,  JHEP {\bf 1404} (2014) 086.

\bibitem{Klevansky:1992qe}
S.~P.~Klevansky, Rev.\ Mod.\ Phys.\  {\bf 64} (1992) 649.

\bibitem{Jansen:2009hr}
K.~Jansen {\it et al.}  [ETM Collaboration], Phys.\ Rev.\ D {\bf 80} (2009) 054510.

\bibitem{Maris:1999nt}
P.~Maris, P.~C.~Tandy, Phys.\ Rev.\ C {\bf 60} (1999) 055214.

\bibitem{Beringer:1900zz}
J.~Beringer {\it et al.}  [Particle Data Group Collaboration], Phys.\ Rev.\ D {\bf 86} (2012) 010001.

\bibitem{Becirevic:2003pn}
D.~Becirevic, V.~Lubicz, F.~Mescia, C.~Tarantino, JHEP {\bf 0305} (2003) 007.

\end{thebibliography}
\end{document}